\documentclass[12pt,preprint]{aastex}
\usepackage{graphicx}

\usepackage{pdflscape}

\slugcomment{The Astronomical Journal}

\shorttitle{Far Out Bodies}
\shortauthors{Jewitt}

\begin{document}

\title{Color Systematics of Comets and Related Bodies\footnote
{The data presented herein were obtained at the W.M. Keck Observatory, which is operated as a scientific partnership amongst the California Institute of Technology, the University of California and the National Aeronautics and Space Administration. The Observatory was made possible by the generous financial support of the W.M. Keck Foundation. 
}   
}

\author{David Jewitt$^{2,3}$}
\affil{$^2$Department of Earth, Planetary and Space Sciences, UCLA, \\
595 Charles Young Drive East, 
Los Angeles, CA 90095-1567\\ 
$^3$Department of Physics and Astronomy, UCLA, \\
430 Portola Plaza, Box 951547
Los Angeles, CA 90095-1547 
}

\email{jewitt@ucla.edu}

\begin{abstract}
Most comets are volatile-rich bodies that have recently entered the inner solar system following long-term storage in the Kuiper belt and the Oort cloud reservoirs.  These reservoirs feed several distinct, short-lived ``small body'' populations.  Here, we present new measurements of the optical colors of cometary and comet-related bodies including long-period (Oort cloud) comets, Damocloids (probable inactive nuclei of long-period comets) and Centaurs (recent escapees from the Kuiper belt and precursors to the Jupiter family comets).  We combine the new measurements with published data on short-period comets, Jovian Trojans and Kuiper belt objects to examine the color systematics of the comet-related populations.  We find that the mean optical colors of the dust in short-period and long-period comets are identical within the uncertainties of measurement, as are the colors of the dust and of the underlying nuclei.  These populations show no evidence for scattering by optically-small particles or for compositional gradients, even at the largest  distances from the Sun, and no evidence for  ultrared matter.  Consistent with earlier work, ultrared surfaces are common in the Kuiper belt and on the Centaurs, but not in other small body populations, suggesting that this material is hidden or destroyed upon entry to the inner solar system. The onset of activity in the Centaurs and the disappearance of the ultrared matter in this population begin at about the same perihelion distance ($\sim$10 AU), suggesting that the two are related. Blanketing of primordial surface materials by the fallback of sub-orbital ejecta, for which we calculate a very short timescale, is the likely mechanism.  The same process should operate on any mass-losing body, explaining the absence of ultrared surface material in the entire comet population.

\end{abstract}


\section{Introduction}

\begin{figure}
\epsscale{0.8}
\begin{center}
\plotone{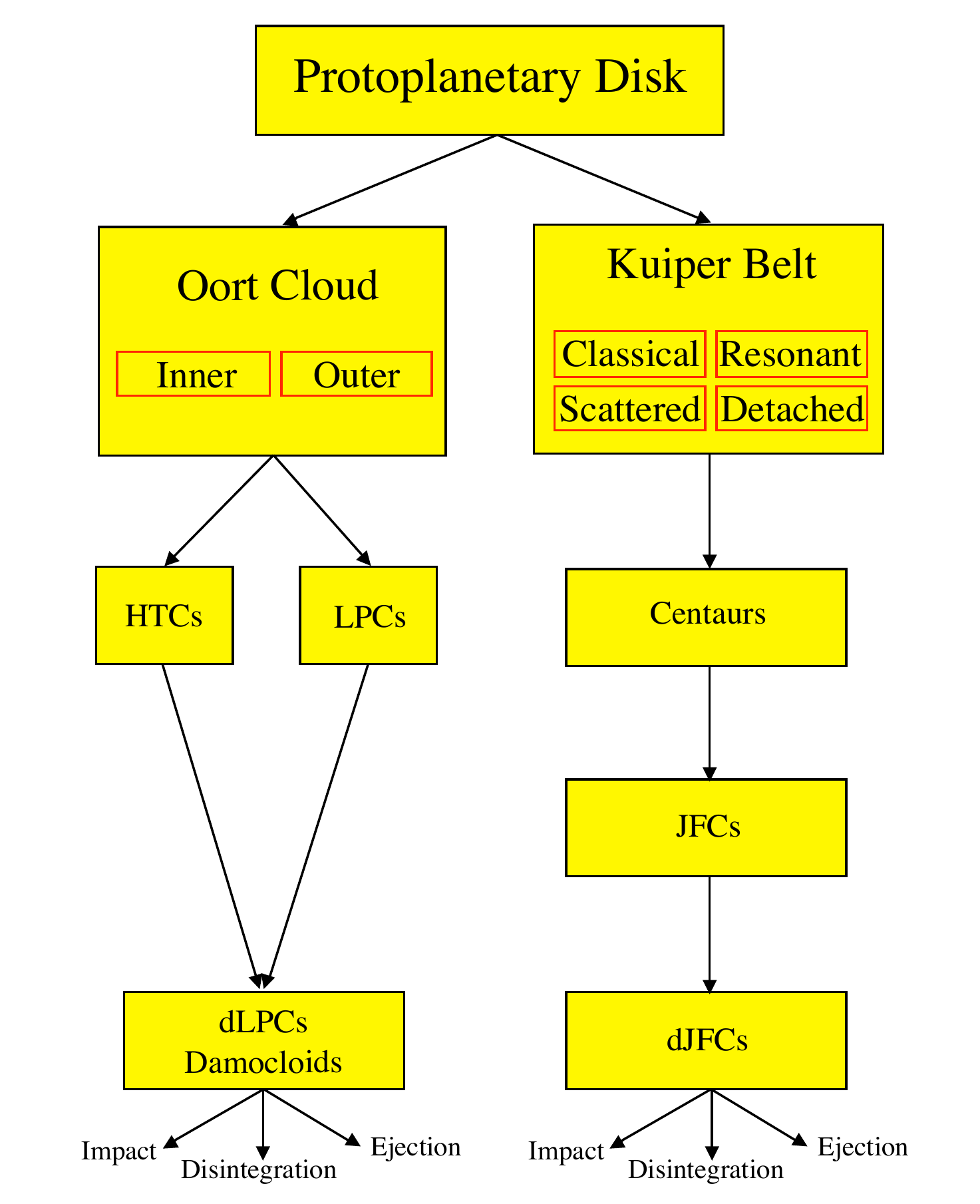}
\caption{Schematic flow diagram highlighting suspected connections between cometary populations. Numbers give the approximate lifetimes of the different stages.  SPC = short-period comet, LPC = long period comet, HTC = Halley-type comet, dLPC and dJFC are defunct LPCs and JFCs, respectively.  Several suggested connections (e.g.~between the Scattered KBOs and the Oort cloud) have been omitted for clarity.  \label{flow}
} 
\end{center} 
\end{figure}

Several  distinct and apparently unrelated solar system populations  are now recognized to be evolutionary states derived mainly from two source reservoirs, the Oort cloud and the Kuiper belt.  These icy reservoirs are recognized as products of accretion in the Sun's long-gone protoplanetary disk, while the details of their relation to this disk are the uncertain subjects of on-going research. Sublimation of ice in Oort cloud and Kuiper belt escapees begins  when their perihelia are displaced towards the Sun, and becomes especially strong inside the orbit of Jupiter ($\sim$5 AU), where crystalline water ice is thermodynamically unstable. The resulting activity leads to the conventional re-labeling of these objects as comets.   Classically, objects are labelled as either short-period comets (SPCs) or long-period comets (LPCs) depending on whether their orbital periods are less than or greater than 200 yr.  This somewhat arbitrary distinction provides a rather good discriminant between comets originating in the two reservoirs.  Broadly speaking,  the Kuiper belt (where equilibrium temperatures are $T \sim$ 30 to 40  K) feeds the population of SPCs while the Oort cloud ($T \sim$ 10 K) is the source of LPCs.  The schematic Figure (\ref{flow}), included here as a guide to the rest of the paper,  shows some of the suggested relationships between the reservoirs and the comet-related populations.  While the Figure is highly simplistic and hides many uncertainties about possible dynamical interrelations (e.g.~Fern{\'a}ndez et al.~2004, Emelyanenko et al.~2013), it serves our purposes here by usefully connecting the different small-body populations discussed in this paper.

The nomenclature used to describe different dynamical types of comet has recently become much more complicated and deserves a brief account.  The comets are sometimes classified by $T_J$, their Tisserand parameters measured with respect to Jupiter:

\begin{equation}
T_J = \frac{a_J}{a} + 2 \left[(1 - e^2)\frac{a}{a_J}\right]^{1/2} \cos(i)
\end{equation}

\noindent where $a$, $e$ and $i$ are the orbital semimajor axis, eccentricity and inclination of the orbit and $a_J$ is the semimajor axis of Jupiter.  This parameter is conserved in the circular, restricted three-body approximation to cometary dynamics.  Jupiter family comets (JFCs), a sub-set of the SPCs, have $2 \le T_J \le 3$ while LPCs have $T_J <$ 2 (Vaghi 1973).  An intermediate group called Halley type comets (HTCs), also with $T_J < 2$, has an inclination distribution distinct from the disk-like JFCs and the isotropically distributed LPCs.  
In this paper we are not particularly concerned with  minute dynamical distinctions between comet classes and, for the most part, are content to distinguish mainly between SPCs and LPCs as originally defined.  We use ``JFC'' and ``SPC'' interchangeably.  Likewise, we use ``Halley type'' and ``long-period'' comets interchangeably.   The terms ``ecliptic comets" and ``nearly isotropic comets" have been suggested (Levison 1996), but these are practically degenerate with the more familiar short-period and long-period appellations and we do not use them here.  

The classical Oort cloud is a 50,000 AU scale structure that supplies LPCs and, probably, HTCs (although the source of HTCs has not been definitively established; c.f.~Levison et al.~2006, Emelyanenko et al.~2013).  Damocloids are point-like objects with $T_J < 2$ that are likely to be the defunct nuclei of HTCs (Jewitt 2005), but again the details of this connection are not certain (Wang et al.~2012).    Objects with perihelia between the orbits of Jupiter and Neptune are called Centaurs. They are likely escaped Kuiper belt objects which, in turn, feed the population of JFCs.  The Centaurs are short-lived (median lifetime $\sim$10 Myr, albeit with a wide range; Horner 2003, 2004, Tiscareno and Malhotra 2003) as a result of strong gravitational scattering interactions with the giant planets.   The JFCs are even shorter-lived ($\sim$0.4 Myr; Levison and Duncan 1994) because of frequent scattering by the terrestrial planets. They also experience very rapid physical evolution driven by sublimation.  About 80\% of the ``asteroids'' possessing JFC-like orbits have low, comet-like albedos (Kim et al.~2014).

Previous work has established that the Kuiper belt objects display an extraordinarily large range of optical colors (Luu and Jewitt 1996, Jewitt and Luu 2001, Tegler and Romanishin 2000, Jewitt 2002, Hainaut and Delsanti 2002, Tegler et al.~2003, Peixinho et al.~2004, Hainaut et al.~2012). This is due, in part, to the presence of ultrared matter which is defined as having a normalized  reflectivity gradient $S' \ge$ 25\% (1000\AA)$^{-1}$, corresponding to optical colors B-R $\ge$ 1.6 magnitudes (Jewitt 2002).  While irradiated organics have long been suspected to be responsible for the red colors, the nature of the ultrared matter remains unknown. We do know that the distribution of ultrared matter in the solar system is peculiar.   It is most concentrated in the dynamically cold (low inclination) portion of the Classical Kuiper belt (Tegler and Romanishin 2000, Trujillo and Brown 2002) but is present in all the known Kuiper belt populations (e.g. Sheppard 2010, Hainaut et al.~2012).  Ultrared matter exists on the Centaurs, whose optical color distribution appears to be bimodal (Tegler et al.~2003, Peixinho et al.~2003, 2012) but it has not been reported on other small-body populations (Jewitt 2002).  

In this paper, we present new optical measurements and combine them with data from the literature in order to examine  color systematics of the comet-related populations.  We  present the  systematic measurements of the LPCs, as a group, and the first measurements of trans-Jovian LPCs (i.e.~those with perihelia beyond Jupiter's orbit, where sublimation of crystalline water ice is negligible).  Our measurements of Damocloids and Centaurs extend earlier work (Jewitt 2005 and 2009, respectively).    The paper is divided as follows; Section 2 describes the observational methods and object samples, Section 3 presents results that are discussed in Section 4, while Section 5 presents a summary.

\section{Observations} 

\subsection{Methods}
We used the 10-meter diameter Keck I telescope located atop Mauna Kea, Hawaii and the Low Resolution Imaging Spectrometer (LRIS) camera (Oke et al.~1995) to obtain photometry.  The LRIS camera has two channels housing red and blue optimized charge-coupled devices (CCDs) separated by a dichroic filter (we used the ``460'' dichroic, which has 50\% transmission at 4875\AA).  On the blue side we used a broadband B filter (center wavelength $\lambda_c$ = 4369\AA, full width at half maximum (FWHM) $\Delta \lambda$ = 880\AA) and on the red side a V filter ($\lambda_c$ = 5473\AA, $\Delta \lambda$ = 948\AA) and an R filter ($\lambda_c$ = 6417\AA, $\Delta \lambda$ = 1185\AA).  Some objects were also measured in the I filter ($\lambda_c$ = 7600\AA, $\Delta \lambda$ = 1225\AA).  The I-filter measurements are listed in the data-tables of this paper but have not been used  in the subsequent analysis because they are few in number compared to photometry in B, V and R.  All observations used the facility atmospheric dispersion compensator to correct for differential refraction, and the telescope was tracked non-sidereally while autoguiding on fixed stars.   The image scale on both cameras was 0.135\arcsec~pixel$^{-1}$ and the useful field of view approximately 320\arcsec$\times$440\arcsec.  Atmospheric seeing ranged from  $\sim$0.7 to 1.3\arcsec~FWHM and observations were taken only when the sky above Mauna Kea was photometric, as judged in real-time from a photometer at the nearby Canada-France-Hawaii telescope and later by repeated measurements of photometric standard stars.

The data were reduced by subtracting a bias (zero exposure) image and then dividing by a flat field image constructed from integrations taken on a diffusely illuminated spot on the inside of the Keck dome.  The target objects were identified in the flattened images from their positions and distinctive non-sidereal motions.  Photometry was obtained using circular projected apertures tailored to the target and the individual nightly observing conditions, with sky subtraction obtained from a contiguous annulus.    In poor seeing we used appropriately enlarged apertures.   Observations of resolved comets generally used a sky annulus 12\arcsec~in radius and 6\arcsec~in width; the resulting contamination of the sky annulus by cometary dust is judged to be minimal.  Photometric calibration was secured from observations of standard stars from Landolt (1992), always using the same apertures as employed for the target objects, on stars at similar airmass.  We used only Landolt stars having $\pm$0.01 magnitude or better uncertainties and colors close to those of the Sun.  Repeated measurements of the standard stars confirmed the photometric stability of each night at the $\sim \pm$1\% level.

The red-side CCD in LRIS is a physically thick device that is particularly susceptible to ``cosmic ray'' (actually muon and other ionizing particles) contamination.  We examined each image for such contamination and individually removed artifacts by digital interpolation before photometry where possible.  In a few cases, long cosmic ray tracks (caused by energetic particles grazing the CCD) could not be removed and we eliminated these images from further consideration.  Likewise, objects whose photometry was contaminated by  field stars and galaxies were revisited where possible, and ignored from further consideration where not. 
Uncertainties were estimated from the scatter of repeated measurements taken within a single night.   Some objects were observed on more than one night.  In general, we find that the agreement between nights is compatible with the photometric uncertainties estimated nightly.

\subsection{The Observational Samples}

The new measurements presented in this paper refer to the populations of long-period comets, the Damocloids, and the Centaurs.  For each of these populations we present the orbital properties, the geometric circumstances of observation and the photometric results in a series of data tables, as we shortly discuss.  The orbital properties are taken from the NASA JPL Horizons ephemeris service (http://ssd.jpl.nasa.gov/horizons.cgi), which was also used to compute the geometric circumstances for each observation.  Photometric uncertainties on each object, $\sigma$, were computed from $\sigma = (k_0^2 + \sigma_d^2)^{1/2}$, where $\sigma_d$ is the standard error on the mean of repeated measurements. Quantity $k_0$ is a floor to the acceptable uncertainty estimated from the scatter in zero points deduced from observations of photometric standard stars, and from inspection of real-time opacity measurements from the CFHT Skyprobe (http://www.cfht.hawaii.edu/Instruments/Elixir/skyprobe/home.html).  Typically, we found $k_0$ = 0.01 while $\sigma_d$ varied strongly with the brightness of the object, as expected, and with the contaminating effects of field stars and galaxies.  The reported mean colors of the different populations are given as unweighted means of the individual object colors, with the error on the mean computed assuming Gaussian statistics (i.e. the error on the mean is approximately the standard deviation of the population divided by the square root of the number of measurements).  We conservatively elected not to consider the weighted mean color of a population because the weighting gives too much power to the most precise photometry (typically of the brightest objects).   However, in most cases, the unweighted mean and the weighted mean colors of a population are consistent. 

\textbf{Long Period Comets:}
We observed 26 LPCs (i.e.~comets with $T_J <$ 2), 18 of them  with perihelion distance $q > a_J$, where $a_J$ = 5.2 AU is the orbital semimajor axis of Jupiter.  Interest in the properties of these rarely-observed trans-Jovian objects is high, for two reasons.  Firstly, beyond Jupiter, the rate of sublimation of crystalline water ice is negligible, meaning that any observed activity must have another cause (either the sublimation of a more volatile ice, or the action of a different mechanism of ejection).  Secondly, trans-Jovian radiation equilibrium temperatures are so low that ice grains expelled from the nucleus can survive in the coma, whereas  the lifetimes of  ice grains in the inner solar system are strongly curtailed by sublimation.   Together, these effects (a potential change in the physics of ejection and the preferential survival of volatile solids at large distances) are expected to have observable effects on the trans-Jovian LPCs.  

The orbital elements of the observed LPCs are listed in Table (\ref{high_q_orbits}).    Two  objects (2013 AZ60 with $q$ = 7.9 AU and 2013 LD16 with $q$ = 2.545 AU) lack a cometary designation but are included in Table (\ref{high_q_orbits}) because we have observed them to show coma.  Of the 18 trans-Jovian comets, three (namely C/2014 B1 at $q$ = 9.5 AU, C/2010 L3 at $q$ = 9.9 AU and C/2003 A2 at $q$ = 11.4 AU) have perihelia at or beyond the orbit of Saturn.

Table (\ref{comet_geometry}) lists the geometric circumstances of observation for each comet, while the color measurements are presented in Table (\ref{lpc_photometry}).  
The mean colors of the LPCs from our observations are B-V = 0.78$\pm$0.02, V-R = 0.47$\pm$0.02 and R-I = 0.42$\pm$0.03.  Some of the LPCs in our sample are likely making their first passage through the planetary system and may show properties different from comets that have been previously heated (for example, owing to the release of surface material accumulated during 4.5 Gyr of exposure in the Oort cloud).  To this end, we analyzed the pre-perihelion and post-perihelion observations separately, finding B-V = 0.81$\pm$0.02, V-R = 0.47$\pm$0.02 and R-I = 0.40$\pm$0.04 (pre-perihelion) and B-V = 0.75$\pm$0.02, V-R = 0.47$\pm$0.02 and R-I = 0.44$\pm$0.03 (post-perihelion), with 13 comets in each group.   No significant  differences exist between the pre- and post-perihelion colors of the long-period comets.  Neither do the colors show a correlation with the orbital binding energy (taken as the inverse semi-major axis, from Table \ref{high_q_orbits}). We conclude that there is no evidence for color differences that might be associated with the first entry of dynamically new comets into the planetary region. This conclusion is  tempered by intrinsic uncertainties in the orbits followed by comets in the past  (e.g.~Krolikowska and Dybczynski 2013).

 For comparison, the mean colors of six LPCs measured by Solontoi et al.~(2012)  in the Sloan filter system (but transformed to BVRI using the relations given by Ivezic et al.~2007) are B-V = 0.76$\pm$0.01 (6), V-R = 0.43$\pm$0.01, are in reasonable agreement with our data.  The mean colors of five LPCs reported by Meech et al.~(2009) are B-V = 0.687$\pm$0.005, V-R=0.443$\pm$0.003.  While V-R is again in good agreement, the latter B-V color appears bluer by $\sim$0.08 magnitudes than in Solontoi et al.~(2012) or the present work, and this difference is unexplained.   Comet C/2003 A2, the only object observed in common between the Meech et al.~paper  and the present work, has consistent colors (V-R = 0.46$\pm$0.01, R-I = 0.39$\pm$0.01 from Meech et al.~vs.~V-R = 0.47$\pm$0.04, R-I = 0.46$\pm$0.04 here) but was not measured in B-R.   Object 2010 AZ60 was independently observed by Pal et al.~(2015), who found Sloan g' - r' = 0.72$\pm$0.05. When transformed according to the prescription by Jester et al.~(2005), this gives B-V = 0.93$\pm$0.06.  This compares with B-V = 0.82$\pm$0.01 measured here.  We consider this reasonable agreement given the large uncertainties on the former measurement. Pal et al.~did not comment on the cometary nature of 2010 AZ60.

\textbf{Damocloids:}
The Damocloids are point-source objects having $T_J < 2$, where $T_J$ is the Tisserand parameter measured with respect to Jupiter (Jewitt 2005).   The orbital elements of the Damocloids observed here are reported in Table (\ref{damo_orbits}) and the geometric circumstances of observation may be found in Table (\ref{damo_geometry}).   The measured  colors of the Damocloids are listed in Table (\ref{damo_photometry}).  The mean colors from these measurements alone are B-V = 0.80$\pm$0.02, V-R = 0.54$\pm$0.01, R-I = 0.45$\pm$0.03 and B-R = 1.34$\pm$0.02.   Combined with additional measurements  from Table (4) of Jewitt (2005), we obtain mean colors  B-V = 0.80$\pm$0.02, V-R = 0.51$\pm$0.02, R-I = 0.47$\pm$0.02 and B-R = 1.31$\pm$0.02.

Two of the 15 observed Damocloids,  C/2010 DG56 and C/2014 AA52, received cometary designations between their selection for this study and their observation at the Keck.  A third, 2013 LD16, was found by us to be cometary, although it retains an asteroidal designation.   This transformation from inactive to active also occurred in our original study of the Damocloids (Jewitt (2005)), and provides strong evidence that bodies selected as  probable defunct comets on the basis of their distinctive orbits indeed carry near-surface volatiles.  We retain the active Damocloids in our sample and note that the weakness of their comae and tails suggest that the colors of their nuclei are still accurately measured.    The colors of object 342842 (2008 YB3) were independently measured by Sheppard (2010) as B-R = 1.26$\pm$0.01, V-R = 0.46$\pm$0.01, and by Pinilla-Alonso et al.~(2013) as  B-R = 1.32$\pm$0.06, V-R = 0.50$\pm$0.06, in reasonable agreement with the colors measured here B-R = 1.25$\pm$0.02, V-R = 0.51$\pm$0.02.

\textbf{Centaurs:}
To define our Centaur sample (Table \ref{centaur_orbits}), we selected objects having perihelia $q > a_J$ and semimajor axes $|a| < a_N$, ignoring objects in 1:1 resonance with the giant planets (c.f.~Jewitt (2009)).  This is a narrower definition than is employed by some dynamicists, but serves to provide a convenient sample distinct from the Kuiper belt objects at larger semimajor axes and the short-period comets at smaller distances.  We observed 17 Centaurs,  7 of them  active.  The geometric circumstances of observation are given in Table (\ref{centaur_geometry}) while the photometry is given in Table (\ref{centaur_photometry}).  To the new observations in these Tables we add  separate measurements of the colors  of  Centaurs using data from Jewitt (2009) and from Peixinho et al.~(2003) and Peixinho et al.~(2012).    We consider the inactive and active Centaurs separately.  Their colors appear quite different in being, for the inactive Centaurs B-V = 0.93$\pm$0.04 (29), V-R = 0.55$\pm$0.03 (29), R-I = 0.45$\pm$0.02(3) and for the active objects B-V = 0.80$\pm$0.03 (12), V-R = 0.50$\pm$0.03 (13).

\textbf{Active Short-Period Comets:}
We used results from a survey by Solontoi et al.~(2012), who found  mean colors    B-V = 0.80$\pm$0.01 (20), V-R = 0.46$\pm$0.01 (20).  These measurements were taken using the SLOAN broadband filter system and transformed to BVR. The transformation may incur a small uncertainty, probably of order 0.01 magnitudes, in addition to the quoted statistical uncertainty.

\textbf{Active Cometary Nuclei:}
The colors of cometary nuclei  have been reported in Tables (3) and (5) of the compilation by Lamy and Toth~(2009).  The accuracy of many of these colors depends on digital processing to remove near-nucleus coma and several discrepant or anomalous objects exist.  We reject objects (for example, 6P/d'Arrest) having wildly inconsistent colors, and objects for which the B-R color is uncertain by $\sigma >$ 0.1 magnitudes.  The resulting sample has 16 short-period comet nuclei (2 $\le T_J \le$ 3), for which the mean colors and the standard errors on the means are B-V = 0.87$\pm$0.05, V-R = 0.50$\pm$0.03, R-I = 0.46$\pm$0.03, giving B-R = 1.37$\pm$0.08.  

The sample also includes 5 comets with $T_J <$ 2, (i.e.~LPC nuclei) for which the mean colors are B-V=0.77$\pm$0.02, V-R = 0.44$\pm$0.02, R-I = 0.44$\pm$0.02(5) and B-R = 1.22$\pm$0.03.

\textbf{Defunct Short-Period Nuclei}
The colors of the defunct nuclei of short-period comets, sometimes called ACOs (Asteroids in Cometary Orbits), have been measured by several authors.  Alvarez-Candal (2013) found a mean reflectivity gradient $S'$ = 2.8\%~(1000\AA)$^{-1}$ (no uncertainty quoted) from a sample of 94 objects, after eliminating 73 objects ``with behavior similar to S- or V- type asteroids'', leading to a value that is likely artificially blue.  Licandro et al.~(2008) reported spectra of 57 objects and from them derived $S'$ = 4.0 \%~(1000\AA)$^{-1}$ (again with no quoted uncertainty). The latter value of $S'$ would correspond approximately to broadband colors B-V = 0.68, V-R = 0.39.   However, both studies note that there is a trend for the colors to become more red as the Tisserand parameter decreases, consistent with dynamical inferences that the ``comet-like orbits'' are fed from a mix of cometary and asteroid-belt sources.  

\textbf{Jupiter Trojans:}
The Jovian Trojans  have no known association with the Kuiper belt or Oort cloud comet reservoirs but we include them for reference because some models posit an origin by capture from the Kuiper belt (Nesvorny et al.~2013).  We take the mean colors of Jovian Trojans from a large study by Szabo et al.~(2007), who reported B-V = 0.73$\pm$0.08, V-R = 0.45$\pm$0.08, R-I = 0.43$\pm$0.10 (where the quoted uncertainties are the standard deviations, not the errors on the means).    The Szabo et al.~sample is very large (N $\sim$ 300) and, as a result, the standard errors on the mean colors are unphysically  small.  We set the errors on the mean colors equal to $\pm$0.02 magnitudes to reflect the likely presence of systematic errors of this order.  Independent colors determined from a much smaller sample (N = 29), B-V = 0.78$\pm$0.09, V-R = 0.45$\pm$0.05, R-I = 0.40$\pm$0.10 are in agreement with the values by Szabo et al.~(Fornasier et al.~2007).

\textbf{Kuiper Belt Objects:}
The colors of Kuiper belt objects are distinguished by their extraordinary range (Luu and Jewitt 1996), and by the inclusion of some of the reddest material in the solar system (Jewitt 2002).  Their colors have been compiled in  numerous sources (e.g.~Hainaut and Delsanti 2002, Hainaut et al.~2012).  Here, we use the online compilation provided by Tegler, \textit{http://www.physics.nau.edu/$\sim$tegler/research/survey.htm}, which is updated from a series of publications (most from Tegler and Romanishin 2000, Tegler et al.~2003) and has the advantages of uniformity and high quality.  Note that the colors of Kuiper belt objects are employed only to provide context to the new measurements, and our conclusions would not be materially changed by the use of another Kuiper belt dataset.  The  mean colors of Kuiper belt objects are B-V = 0.92$\pm$0.02, V-R = 0.57$\pm$0.02.  It is well known that the KBO colors are diverse, and so we additionally compute colors for dynamical sub-groups (the hot and cold Classical KBOs, the 3:2 resonant Plutinos and the scattered KBOs), again using measurements from Tegler et al.~for uniformity.

\section{Results}

The color data are summarized in Table (\ref{summary}) where, for each color index and object type, we list the median color and the mean value together with the error on the mean and, in parentheses, the number of objects used.   We note an agreeable concordance between the median and mean colors of most object classes,  showing that the color distributions are not highly skewed (the main exception is found in the B-R color of the inactive Centaurs, with median and mean colors differing by nearly three times the error on the mean).  We note that the listed uncertainties are typically $\pm$0.02 magnitudes and larger, and that even the B-V color of the Sun has an uncertainty of 0.02 magnitudes (Holmberg et al.~2006).   

We are first interested to know if the measured colors of the LPCs might be influenced by the angular sizes of the apertures used to extract photometry.  If so, the colors of objects measured using fixed angle apertures might appear to depend on geocentric distance.  Color gradients are expected if, for example, the nuclei and the dust coma of a given object have intrinsically different colors, or if the particles properties change with time since release from the nucleus.  In almost every comet observed here, inspection of the surface brightness profiles shows that the cross-section in the photometry aperture is dominated by dust, not by the geometric cross-section of the central nucleus.    We also note that spectra of distant comets are invariably continuum dominated (indeed, gaseous emission lines are usually not even detected, e.g.~Ivanova et al.~2015) because of the strong inverse distance dependence of resonance-fluorescence.  

\begin{figure}
\epsscale{0.95}
\begin{center}
\plotone{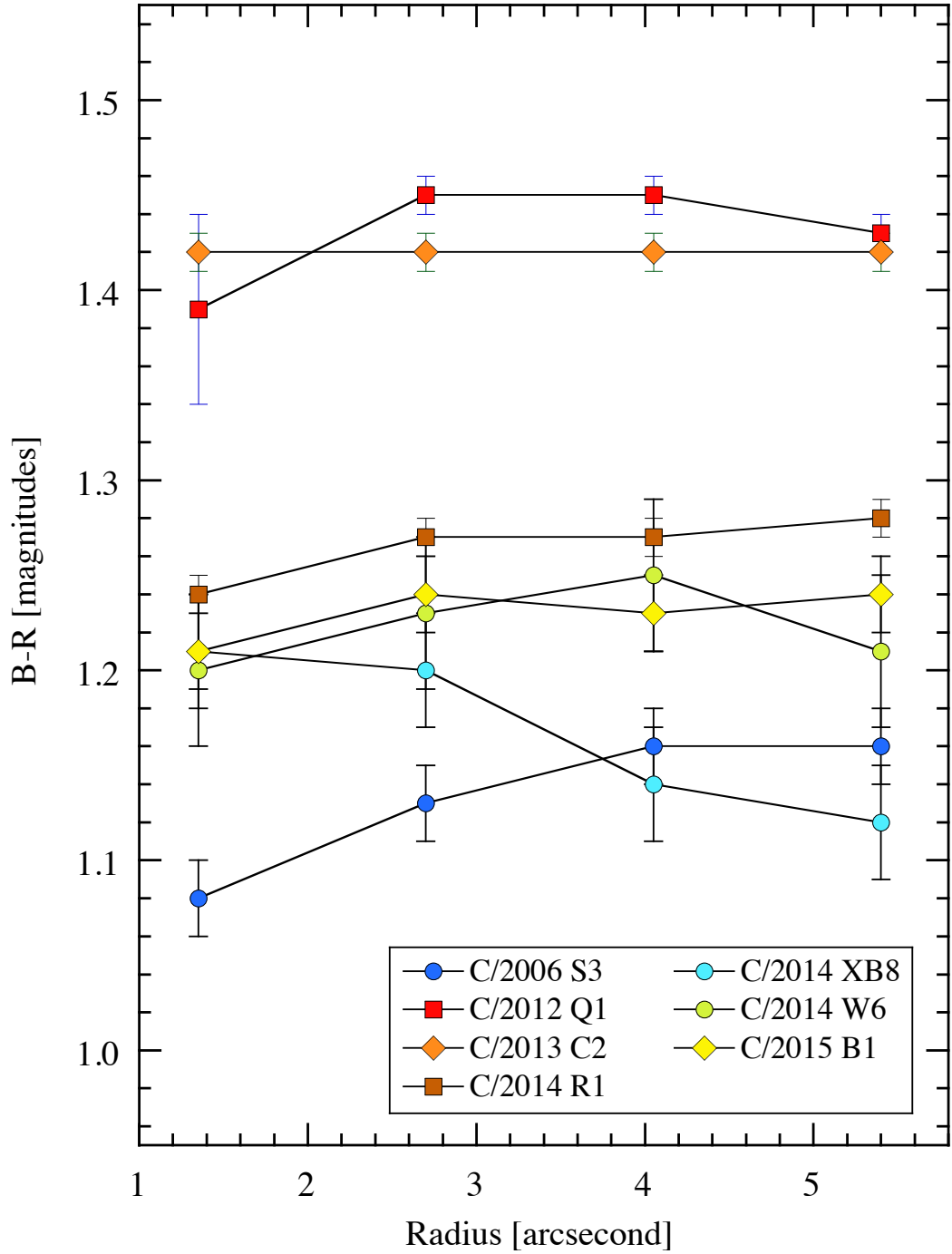}
\caption{The B-R color as a function of projected aperture radius for active comets   \label{BR_vs_radius}
} 
\end{center} 
\end{figure}

We compare measurements of active comets using apertures having different angular radii in Table (\ref{color_gradient}) and show them graphically in Figure (\ref{BR_vs_radius}).  To maximize the color differences and show the results most clearly, we consider only the B-R color index.  The data show that real color variations with angular radius do exist in some active comets but that these variations are small, never larger than 0.1 magnitudes in B-R over the range of radii sampled and more typically only a few $\times$0.01 magnitudes.  Moreover, the color gradients with angular radius can be both positive (as in C/2006 S3) or negative (e.g.~C/2014 XB8).  These findings are consistent with measurements reported for five active Centaurs (Jewitt 2009).  For the comets in Figure (\ref{BR_vs_radius}), radial color variations are  likely to result from small changes in the particle properties as a function of the time since release from the nucleus.  For example, dust fragmentation and/or the loss of embedded volatiles could cause color gradients across the comae.  Here, we merely note the possible existence of these effects and remark on their evident small size.  There is no evidence from the measured comets that the comae are systematically different from the central nuclei.  

\begin{figure}
\epsscale{0.95}
\begin{center}
\plotone{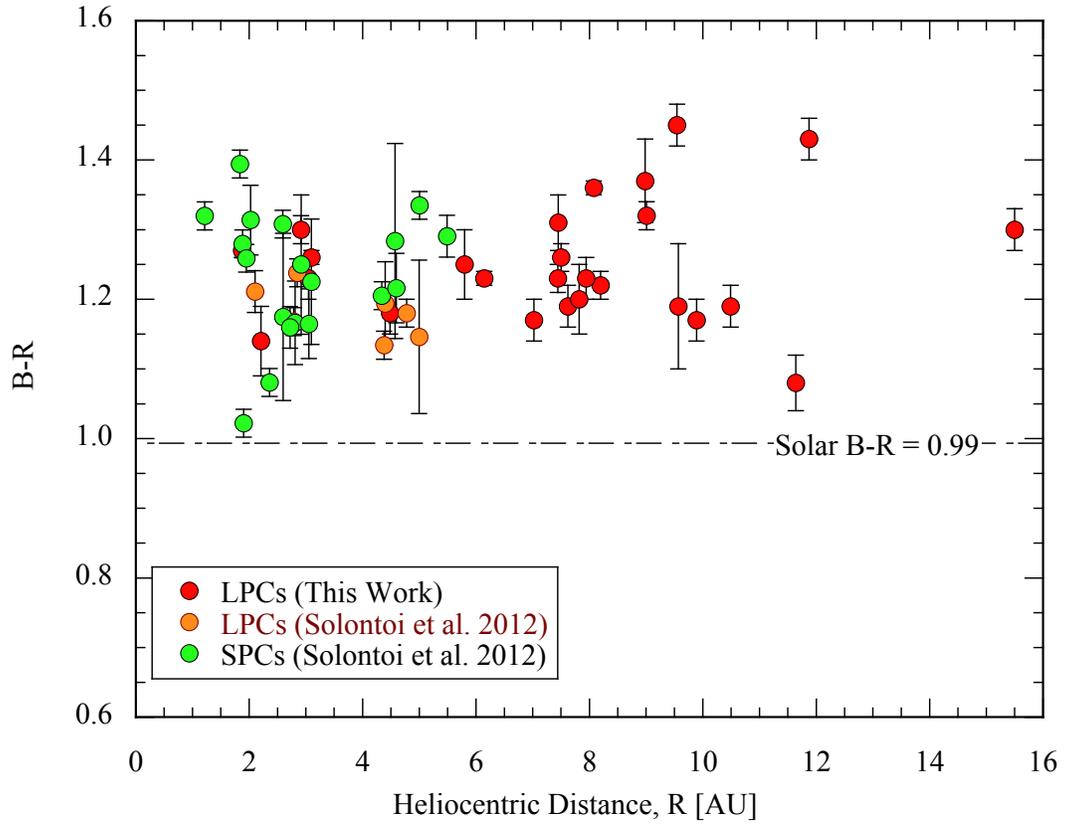}
\caption{B-R color of  LPCs vs. heliocentric distance in AU, from Table (\ref{lpc_photometry}) (red circles) and comets from the SLOAN survey by Solontoi et al.~(2012).  The latter are divided into short period (green circles) and long-period (orange circles) comets.  \label{BR_vs_R}
} 
\end{center} 
\end{figure}

Figure (\ref{BR_vs_R}) shows B-R of the comets as a function of the heliocentric distance, $r_H$, including data from Solontoi et al.~(2012) for comets inside Jupiter's orbit.   The Figure shows no evidence for a trend,  consistent with Jewitt and Meech (1988), Solontoi et al.~(2012) but over a much larger range of heliocentric distances ($\sim$1 AU to $\sim$12 AU) that straddles the water ice sublimation zone near 5 to 6 AU.

\begin{figure}
\epsscale{0.95}
\begin{center}
\plotone{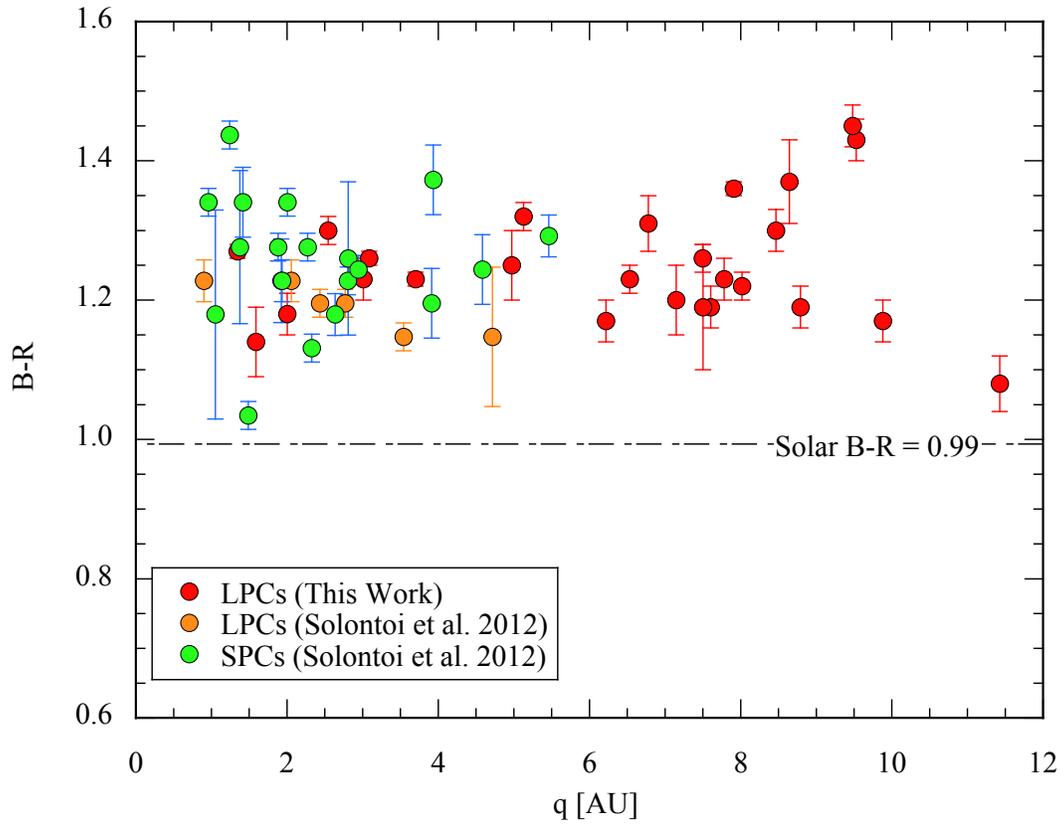}
\caption{B-R color vs.~perihelion distance for LPCs from Table (\ref{lpc_photometry}) (red circles) and comets from the SLOAN survey by Solontoi et al.~(2012).  The latter are divided into short period (green circles) and long-period (orange circles) comets.   No evidence for a color vs.~perihelion distance trend is apparent. \label{BR_vs_q}
} 
\end{center} 
\end{figure}

Figure (\ref{BR_vs_q}) shows the B-R color of the comets as a function of the perihelion distance, $q$.  Clearly, there is no discernible relation between B-R and $q$.   This is significant because, as noted above, the mechanism for mass loss likely changes with increasing distance from the Sun.  Comets active beyond the orbit of Jupiter are likely to be activated by the sublimation of ices more volatile than water and/or by different processes (for example, the exothermic crystallization of ice).  The mean colors of the active short- and long-period comets are indistinguishable in the Figure, mirroring the absence of significant compositional differences between these two groups in measurements of gas-phase species (A'Hearn et al.~2012).

\begin{figure}
\epsscale{0.95}
\begin{center}
\plotone{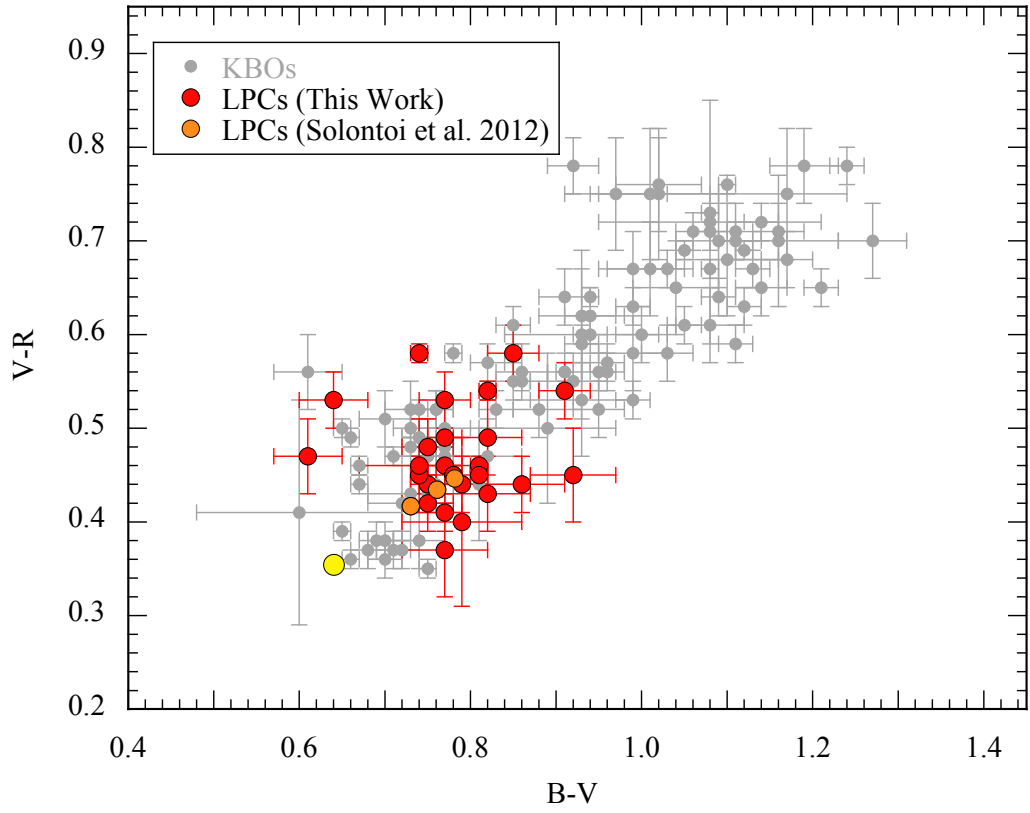}
\caption{Color-color plot comparing LPCs from Table (\ref{lpc_photometry}) (red circles) with Kuiper belt objects (grey circles) measured by Tegler.  The color of the Sun is marked by a yellow circle.   \label{vr_vs_bv}
} 
\end{center} 
\end{figure}

The LPC colors are compared with the colors of Kuiper belt objects  in Figure (\ref{vr_vs_bv}).  The color of the Sun (Holmberg et al.~2006) is  shown as a yellow circle.   While both sets of objects fall on the same color-color line, the comets show no evidence for redder colors (B-R $\ge$ 1.50, B-V $\ge$ 0.95) that are commonly found in members of the Kuiper belt.  
Figure (\ref{vr_vs_bv}) shows that the LPCs are devoid of the ultrared matter.  The significance of the difference is self-evident from the Figure.  It can be simply quantified by noting that none of the 26 measured LPCs has colors redder than the median color of cold classical KBOs.  The probability of finding this asymmetric distribution by chance is (1/2)$^{26} \sim$ 10$^{-8}$, by the definition of the median.   The Kolmogorov-Smirnov (KS) test applied to the color data shows that the likelihood that the B-R distributions of the LPCs and KBOs are drawn from the same parent distribution is $\le$0.001, confirming a $>$3$\sigma$ difference.

\begin{figure}
\epsscale{0.95}
\begin{center}
\plotone{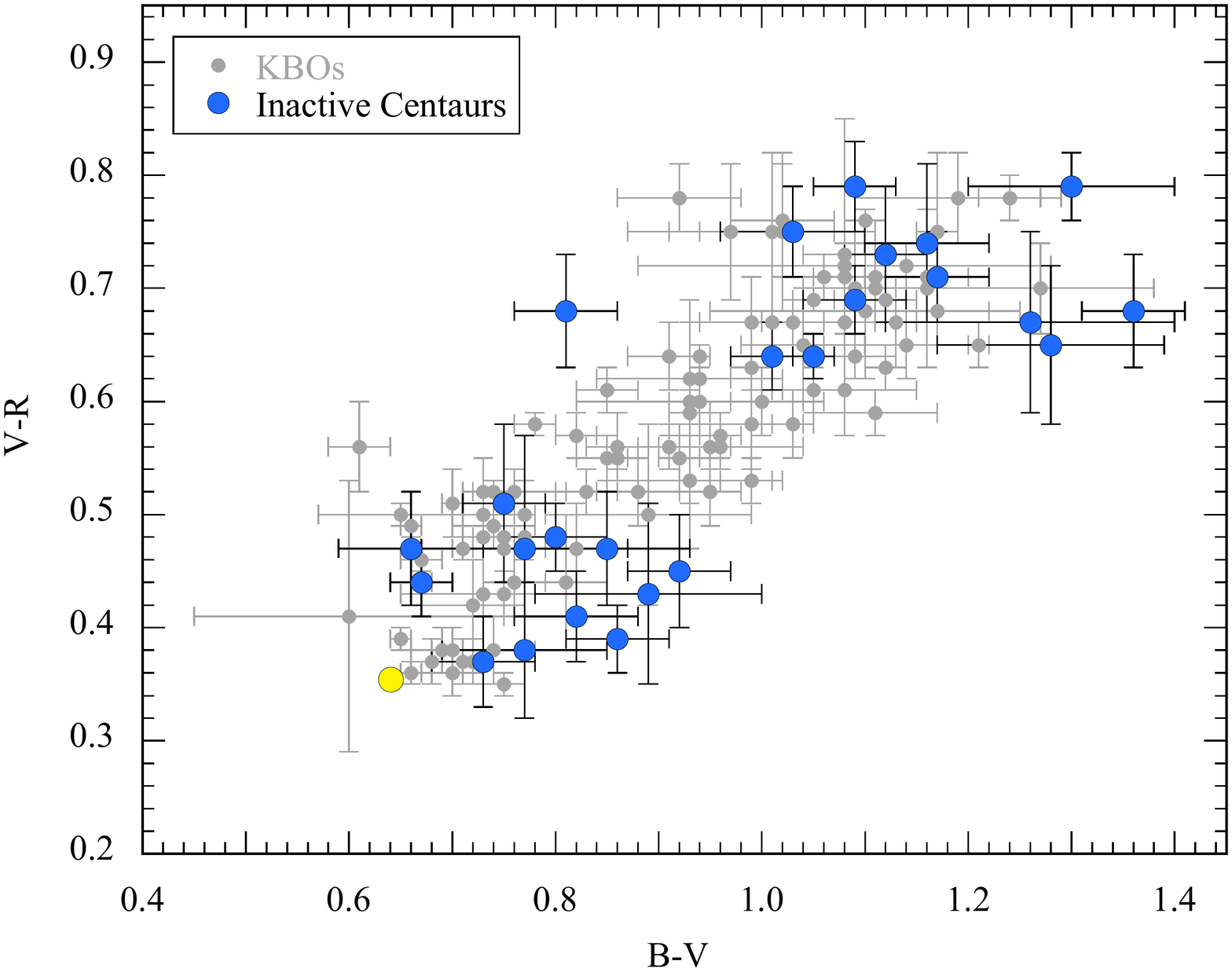}
\caption{Color-color diagram comparing the inactive Centaurs (blue) with Kuiper belt objects (grey).    The yellow circle shows the color of the Sun.  \label{centaurs_1}
} 
\end{center} 
\end{figure}

The colors of the  inactive (point-source) Centaurs (blue circles) are plotted with their formal uncertainties in Figure (\ref{centaurs_1}), where they are compared with the color distribution of the Kuiper belt objects (grey circles) taken from Tegler.  The mean colors of the inactive Centaurs are B-V = 0.93$\pm$0.04 (29), V-R = 0.55$\pm$0.03 (29) and B-R = 1.47$\pm$0.06 (N = 32), overlapping the mean colors of the KBOs (B-V = 0.91$\pm$0.02, V-R = 0.57$\pm$0.01 and B-R = 1.48$\pm$0.03).   Not only do the mean colors agree, but Figure (\ref{centaurs_1}) shows that the inactive Centaurs and the KBOs span the same large range of colors.  The inactive Centaurs, however, show B-V, V-R colors that are bimodally clustered towards the ends of the color distribution defined by the KBOs, with few examples near the mean color of the KBOs.  This bimodality was noted by Peixinho (2003), Tegler et al.~(2003) and Peixinho et al.~(2012), and is an unexplained but apparently real feature of the Centaur color distribution.  It indicates that there are two surface types with few intermediate examples.

\begin{figure}
\epsscale{0.95}
\begin{center}
\plotone{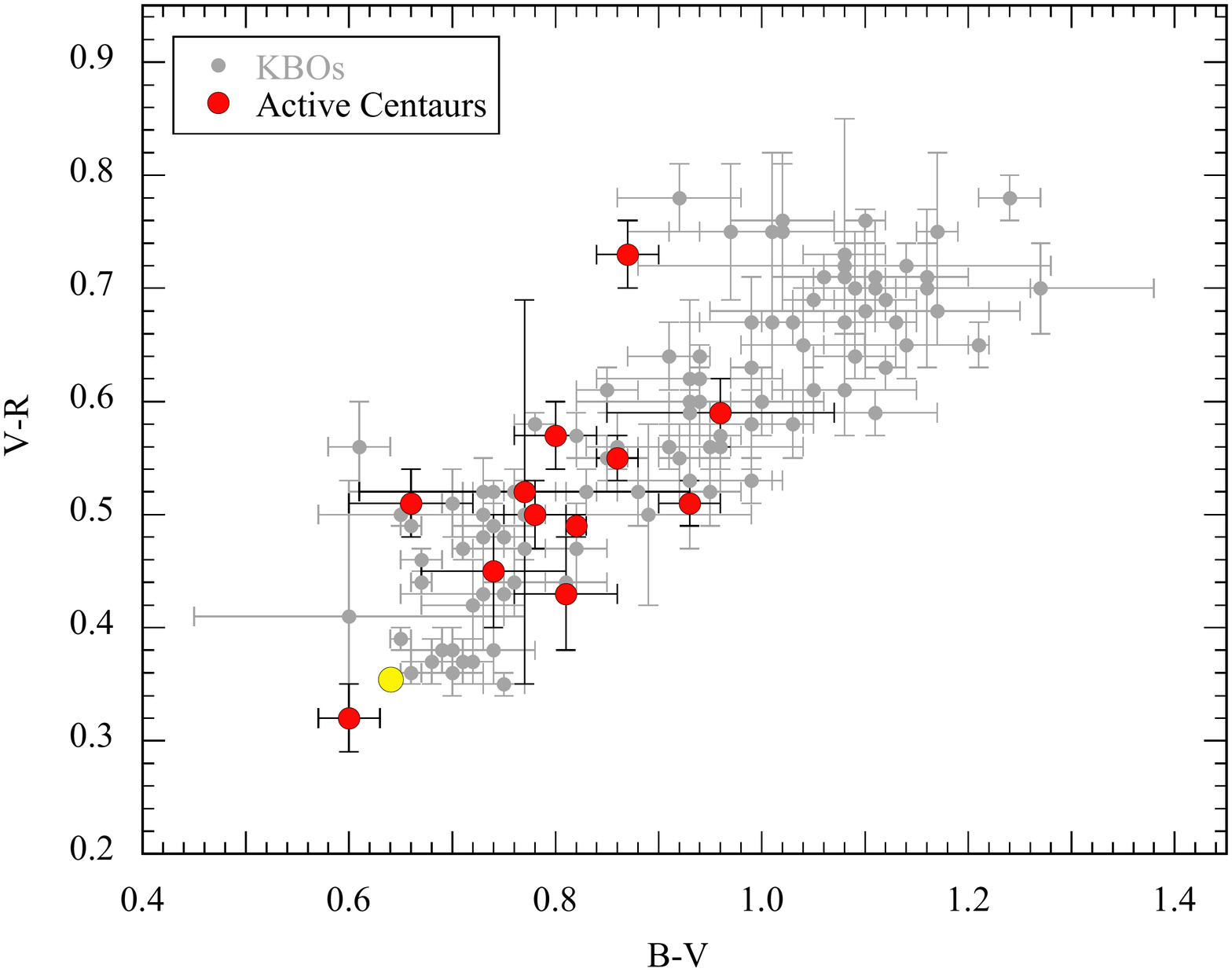}
\caption{Color-color diagram comparing the active Centaurs (red) with Kuiper belt objects (grey).  The yellow circle shows the color of the Sun.  \label{centaurs_2}
} 
\end{center} 
\end{figure}

\begin{figure}
\epsscale{0.8}
\begin{center}
\plotone{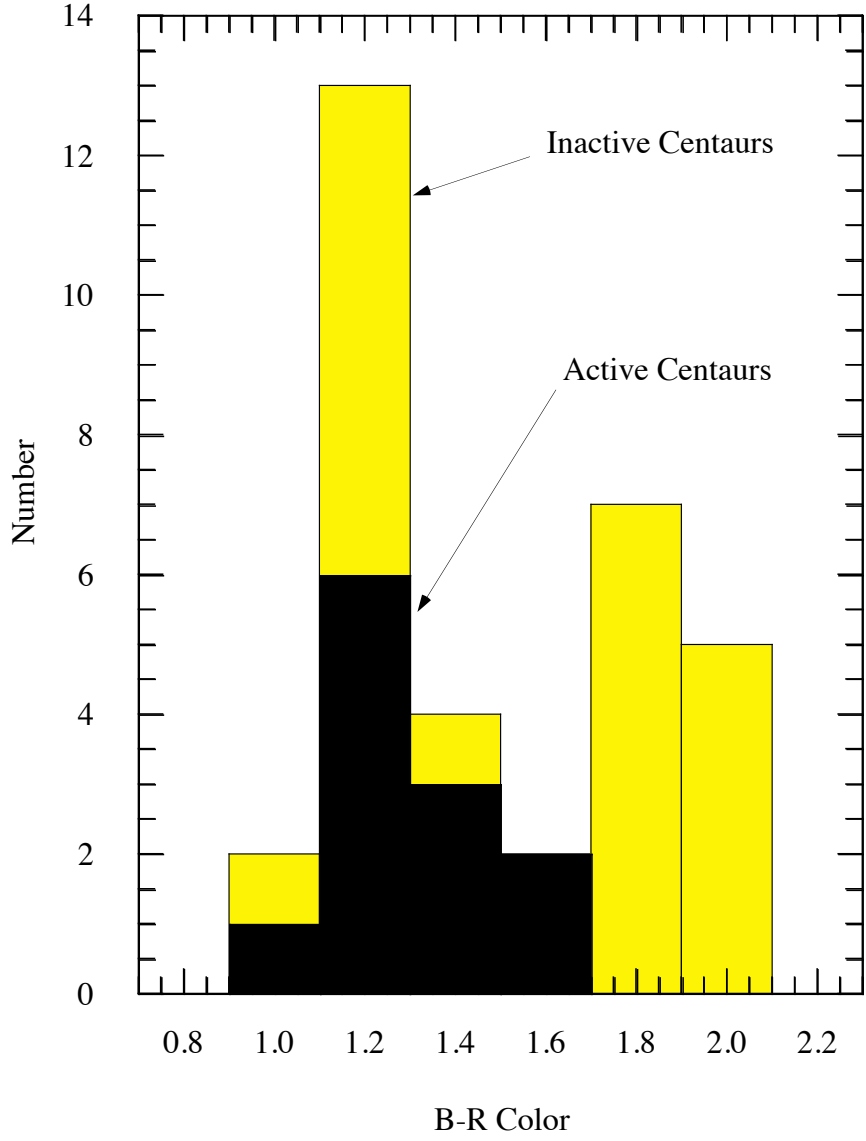}
\caption{Histogram of the B-R colors of active (black) and inactive (yellow) Centaurs. \label{BR_histo}
} 
\end{center} 
\end{figure}

By comparison, the mean colors of the active Centaurs are less red, with B-V = 0.80$\pm$0.03, V-R = 0.50$\pm$0.03, B-R = 1.30$\pm$0.05, while  comparison of Figures (\ref{centaurs_1}) and (\ref{centaurs_2}) shows that the distribution of the colors is qualitatively different from that of the inactive Centaurs. The active Centaur distribution is unimodal, with most of the active Centaurs being  less red than the average color of the KBOs.  The B-R color distributions are compared as a histogram in Figure (\ref{BR_histo}).  

We estimate the statistical significance of the difference between the active Centaurs and the KBOs using the non-parametric Kolmogorov-Smirnov test to compare B-R.  Based on this test, the likelihood that the two populations could be drawn by chance from a single parent population is $\sim$6\%.  In a Gaussian distribution, this would correspond roughly to a 2$\sigma$ (95\%) difference.  Thus, while Figure (\ref{centaurs_2}) is suggestive, we cannot yet formally conclude at the 3$\sigma$ level of confidence that the active and inactive Centaurs have different color distributions.  

\begin{figure}
\epsscale{0.95}
\begin{center}
\plotone{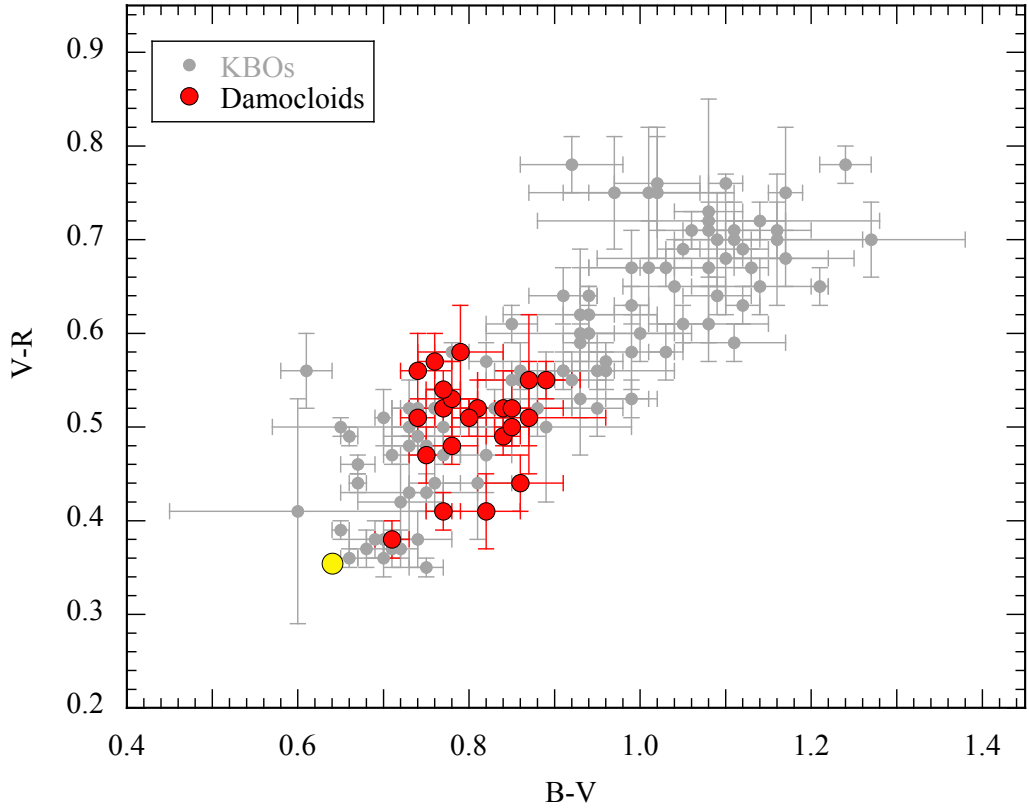}
\caption{Color-color diagram comparing the Damocloids (red circles) with Kuiper belt objects.  The yellow circle shows the color of the Sun.  \label{damos}
} 
\end{center} 
\end{figure}

Figure (\ref{damos}) shows the Damocloids on the Kuiper belt  color field plot.  The new data confirm the absence of ultrared matter previously noted in the Damocloid population (c.f.~Jewitt 2005) but in a sample that is twice as large.  

\section{Discussion}

\begin{figure}
\epsscale{0.85}
\begin{center}
\plotone{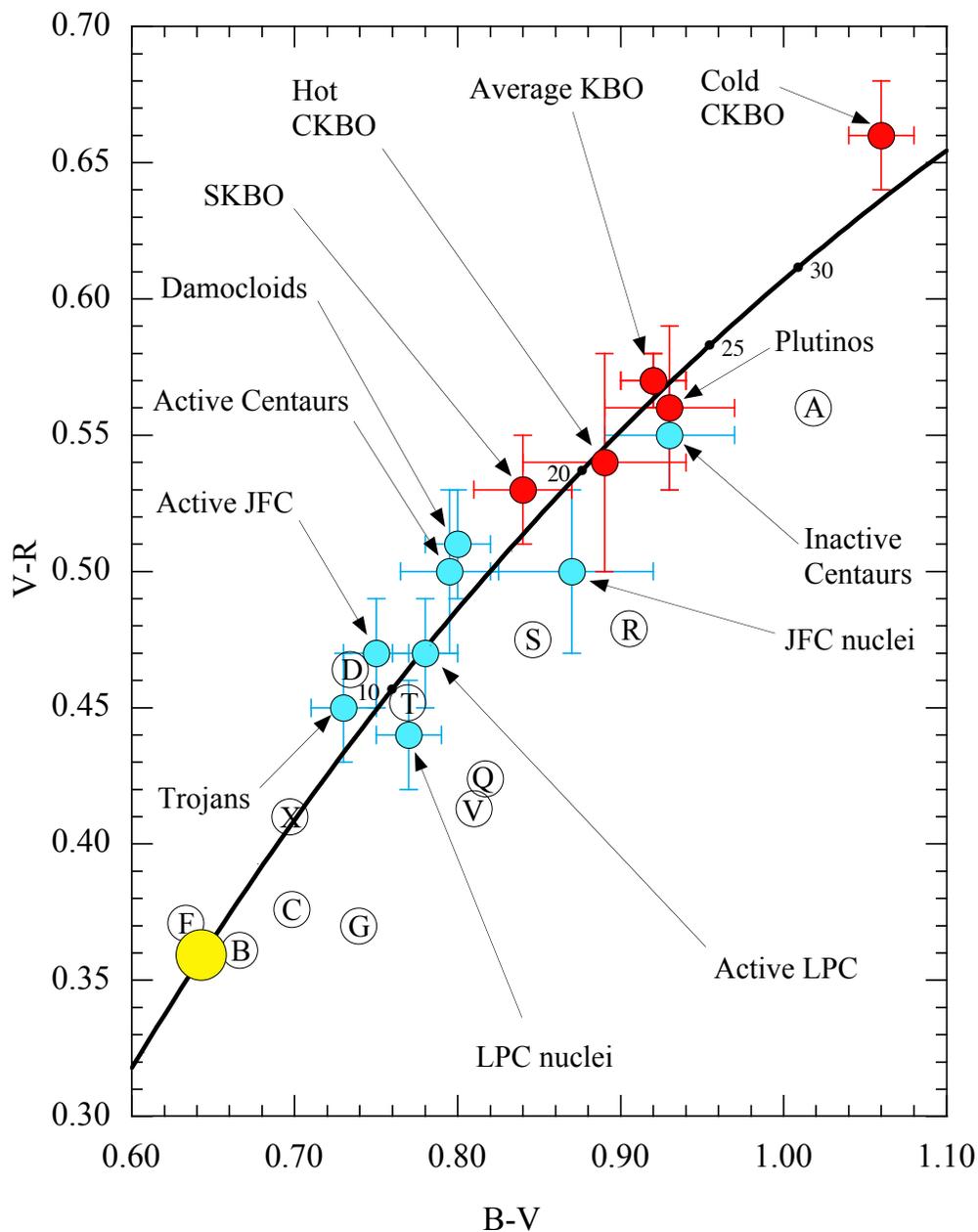}
\caption{Color-color diagram showing the locations of various small-body populations, as labelled. Dynamically distinct subsets of the Kuiper belt are shown as red circles while measurements of comet-related bodies are shown as blue circles (c.f. Table \ref{summary}).  Black circled letters denote asteroid spectral types in the Tholen classification system, from Dandy et al.~(2003).  The solid line shows the locus of points for reflection spectra of constant gradient, $S'$; numbers give the slope in units of \%~1000\AA$^{-1}$. The large yellow circle shows the color of the Sun. Some overlapping points have been displaced (by 0.005 magnitudes) for clarity.   Error bars show the uncertainties on the respective means. \label{color_field}
} 
\end{center} 
\end{figure}

The summary data from Table (\ref{summary}) are plotted in Figure (\ref{color_field}).  Also plotted for reference in Figure (\ref{color_field}) are the colors of common asteroid spectral types, tabulated by Dandy et al.~(2003).   The  solid line in color-color plot  Figure (\ref{color_field}) is the reddening line for objects having linear reflectivity gradients, $S' = dS/d\lambda$ = constant, where $S$ is the normalized ratio of the object brightness to the solar brightness (Jewitt and Meech 1988).  Numbers along the line show the magnitude of $S'$, in units of \% (1000~\AA)$^{-1}$.  Deviations from the line indicate spectral curvature, with objects below the line having concave reflection spectra across the B to R wavelengths ($d^2S/d\lambda^2 > 0$) while those above it are convex ($d^2S/d\lambda^2 < 0$).  The Figure shows that, whereas the reflection spectra of many common asteroid types are concave in the B - R region (as a result of broad absorption bands), the comet-related populations all fall close to the reddening line, consistent with having linear reflectivity spectra.  We emphasize that the reddening line has zero free parameters and is not a fit to the data.  That the mean color measurements fall within a few hundredths of a magnitude from the reddening line gives us confidence that the uncertainties (which are on the same order) have been correctly estimated. However, it is important to remember that these are the average colors of each group, and that individual objects can have colors and spectral curvatures widely different from the average values.

The comet-related populations, including the active comets themselves, are systematically redder than the C-complex asteroids (C, B, F, G) that are abundant members of the outer asteroid belt.  The Jovian Trojans, the least-red of the plotted objects, have mean optical colors that are identical to those of the D-type asteroids within the uncertainties of measurement.  The most red objects in Figure (\ref{color_field}) are the low-inclination ($i \le 2\degr$) Classical KBOs. 

\subsection{The Absence of Blue Scattering in Cometary Dust}
\label{noblue}

We find no significant difference between the mean optical colors of dust in short-period (Kuiper belt) and long-period (Oort cloud) comets.  Comparable uniformity is observed in  the gas phase compositions of comets, where significant abundance differences exist but  are uncorrelated with the dynamical type (Cochran et al.~2015). 
In both dust and gas, relative uniformity of the properties is consistent with radial mixing of the source materials in the protoplanetary disk, and with population of the Kuiper belt and Oort cloud from overlapping regions of this disk.  

We expect that the colors of cometary dust should become more blue with increasing heliocentric distance, for two reasons.  First, the size of the largest particle that can be accelerated to the nucleus escape velocity is a strong function of the gas flow and hence of the heliocentric distance.  At large distances the mean size of the ejected particles should fall into the optically small regime (defined for a sphere of radius $a$ by $x < 1$, where $x = 2\pi a/\lambda$ is the ratio of the particle circumference to the wavelength of observation),  leading to non-geometric scattering effects that include anomalous (blue) colors (Bohren and Huffman 1983, Brown 2014).  Second, ice grains have been detected spectrally in comets (e.g.~Kawakita et al.~2004, Yang et al.~2009, 2014) and should be more abundant in distant comets as a result of  their lower temperatures, reduced sublimation rates and longer lifetimes.  Pure ice grains are bluer than more refractory silicate and carbonaceous solids and thus  the average color should again become more blue with increasing distance from the Sun.  Indeed, Hartmann and Cruikshank (1984) reported a color-distance gradient in comets but this has not been independently confirmed (Jewitt and Meech 1987, Solontoi et al.~2012) and is absent in the present work (Figure \ref{BR_vs_R}).  While one might expect that  the dust in distant comets should  begin to show blue colors consistent with small-particle scattering and with an increasing ice fraction, this is not observed.   We briefly discuss why this might be so.

With $\lambda$ = 0.5 $\mu$m, $x \sim 1$ corresponds to particle radii $a_c \sim$ 0.1 $\mu$m.  Larger particles scatter in proportion to their physical cross-section while in smaller particles the scattering cross-section also depends on the wavelength. The limiting case is Rayleigh scattering, reached in the limit $x \ll$ 1.  The absence of a color-distance trend in comets is most simply interpreted as evidence that the weighted mean particle size, $\overline{a}$, satisfies $\overline{a} > a_c$.   The scattered intensity from a collection of particles is weighted by the size distribution $n(a)da$ (equal to the number of dust grains having radii in the range $a$ to $a + da$), the individual cross sections  ($\propto Q(a) a^2$), where $Q(a)$ is the dimensionless scattering efficiency (c.f.~Bohren and Huffman 1983) and to the residence time, $t$, in the aperture used to measure the brightness.  The latter depends inversely on the speed of the ejected particles, $t \propto v^{-1}$.   Under gas drag acceleration, $v \propto a^{-1/2}$ leading to $t \propto a^{1/2}$. Larger, slower particles spend longer in the photometry aperture than smaller, faster particles and so are numerically over-represented in proportion to $a^{1/2}$.

Accordingly, we write the weighted mean particle size (Jewitt et al.~2014)

\begin{equation}
\overline{a} = \frac{\int_{a_{0}}^{a_{1}} a Q(a) \pi a^{5/2} n(a)da}{\int_{a_{0}}^{a_{1}}  Q(a) \pi a^{5/2} n(a)da}
\label{abar}
\end{equation}

\noindent in which $a_{0}$ and $a_{1}$ are the minimum and maximum sizes in the distribution.  The exact form of $Q(a)$ depends not only on the size, but on the shape, porosity and complex refractive index of the particle, none of which are known from observations.  The limiting cases are $Q(a) = 1$ for $x \gg 1$ and $Q(a) \sim x^{-4}$ (Rayleigh scattering) for $x \ll 1$.

\begin{figure}
\epsscale{0.9}
\begin{center}
\plotone{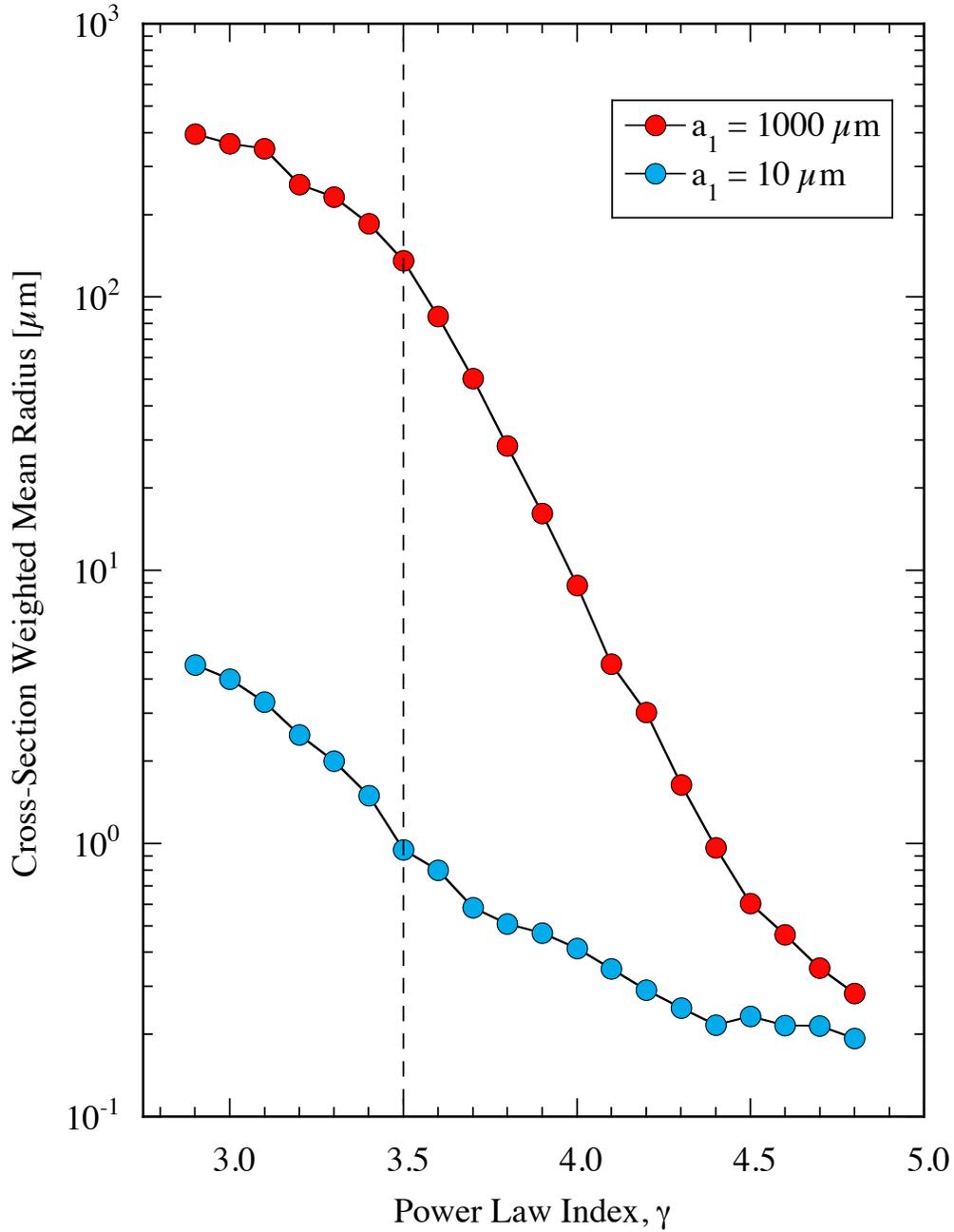}
\caption{Cross-section and residence-time weighted mean particle radius as a function of the size distribution power law index, $\gamma$, computed from Equation (\ref{abar}) (see discussion in section \ref{noblue}). The vertical dashed line marks $\gamma$ = 3.5, a nominal value measured in comets.  \label{figure_abar}
} 
\end{center} 
\end{figure}

Two illustrative solutions to Equation (\ref{abar}) are shown in Figure (\ref{figure_abar}).  For both, we  represent the size distribution by $n(a)da = \Gamma a^{-\gamma} da$, with $\Gamma$ and $\gamma$ constants.   Detailed measurements show size-dependent deviations from power-law behavior, but measured indices are typically in the range $3 < \gamma < 4$, with an average near $\gamma$ = 3.5, which we take as our nominal value (e.g.~Grun et al.~2001, Pozuelos et al.~2014).  We assume that $Q$ = 1 for $a \ge$ 0.1 $\mu$m and $Q = x^4$ otherwise.  We set $a_{0}$ = 10$^{-9}$ m based on measurements from impact detectors on spacecraft near $r_H \sim$ 1 AU (e.g.~Horz et al.~2006).  Smaller particles may exist (although 10$^{-9}$ m is already approaching the dimensions of a big molecule) but the choice of $a_0$ is not critical because $Q(a) \rightarrow$ 0 as $a \rightarrow$ 0.  We evaluate two cases, for $a_1$ = 10 $\mu$m and 1000 $\mu$m, to illustrate the effect of the upper size limit, which is itself set by the rate of sublimation and therefore by the heliocentric distance.  In equilibrium sublimation of exposed, perfectly absorbing water ice, for example, 10 $\mu$m particles can be ejected against gravity by gas drag from a 2 km radius nucleus out to $r_H \sim$ 4.6 AU while 1000 $\mu$m particles are ejected out to $r_H \sim$ 3 AU.

The Figure shows that, when the inefficiency of small particle scattering is included, the inequality $\overline{a} < a_c$ is never reached.  Physically, this is because the optically small particles, although very numerous, are not abundant enough to dominate the scattering cross-section, which therefore reflects optically large, $x > 1$, particles.  This result has been reached from impact counter measurements on spacecraft flying through the comae of active comets (Kolokolova et al.~2004 and references therein).  The Figure shows that the result is generally true, even in the most distant, least active comets.

The role of ice grains in determining the colors of comets also seems to be limited.  One notable exception is provided by 17P/Holmes, which displayed a blue reflection spectrum in the near infrared (1 $\le \lambda \le$ 2.4 $\mu$m) due to small ice grains (Yang et al.~2009).  The exceptional behavior of this comet is likely a transient consequence of its massive outburst, which released abundant, but short-lived ice grains from cold regions beneath the nucleus surface.  Even in 17P/Holmes, the Yang et al.~water ice absorption bands have depths equal to only a few percent of the local continuum, showing dilution of the scattered blue light from small particles by spectrally bland (non-ice) coma dust. 

In the simplest, ``classical'' model (whose origin lies with Whipple's epochal 1950 paper) the largest  particle that can be lifted from the surface is defined by setting the gas drag force equal to the gravitational force towards the nucleus.  In this model, with gas supplied by the sublimation of crystalline water ice, activity is confined to distances less than or comparable to the $\sim$5 AU radius Jupiter's orbit.  Activity in comets with perihelia $\gtrsim$5 AU requires the sublimation of a more volatile ice (e.g.~CO$_2$ or CO) or the action of another process (e.g.~exothermic crystallization of amorphous water ice).

However, many authors (most recently Gundlach et al.~2015) point out that the classical model incorrectly neglects the  effects of particle cohesion, which should be particularly effective in binding small particles to the nucleus surface. In their model, the size distribution of escaping grains is biased toward larger sizes, because of cohesive forces.  This is qualitatively consistent with the absence of color evidence for blue colors and optically small particles in comets.  However,  their ``sticky particle'' model (see their Figure 3) predicts that cohesive forces are so strong that no particles of any size can be ejected by water ice sublimation beyond $r_H \sim$ 2.5 AU and even the more effusive sublimation of super-volatile carbon monoxide ice cannot eject  grains beyond $r_H \sim$ 5 AU.  The activity observed in the distant comets of Table (\ref{comet_geometry}) is thus entirely unexplained.

\subsection{Active Centaurs}

\begin{figure}
\epsscale{1.0}
\begin{center}
\plotone{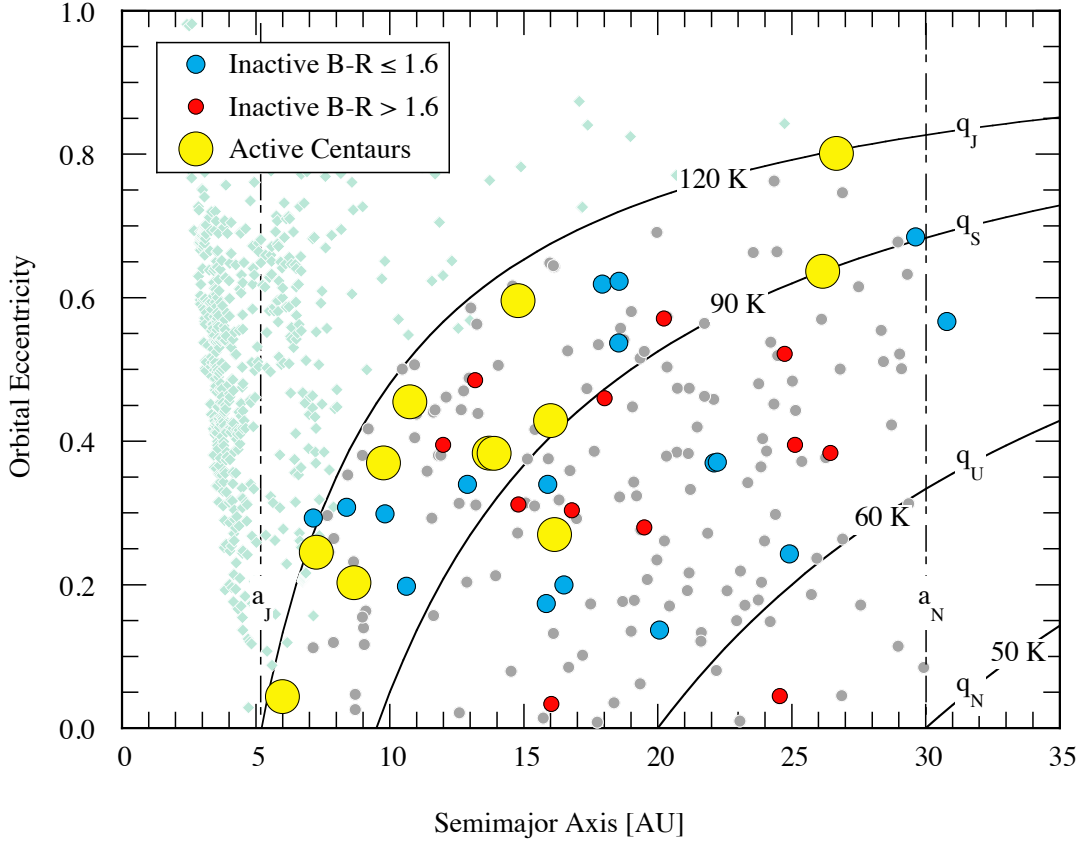}
\caption{Orbital semimajor axis vs.~eccentricity for active Centaurs (large yellow circles), inactive Centaurs with measured B-R $\le$ 1.6 (small blue circles) and inactive Centaurs with B-R $>$ 1.6 (small red circles). Also plotted are all known Jupiter family comets (ocean green diamonds) and Centaurs (grey circles).  Diagonal arcs mark the locus of orbits having perihelion distances equal to the orbital semimajor axes of the giant planets (labeled $q_J$, $q_S$, $q_U$ and $q_N$ for Jupiter, Saturn, Uranus and Neptune, respectively).  The spherical blackbody temperature for each perihelion distance, $T_{BB}$,  is marked.  Vertical dashed lines denote the semimajor axes of Jupiter ($a_J$) and Neptune ($a_N$), for reference. \label{a_e_colors_nolabels}
} 
\end{center} 
\end{figure}

Figure (\ref{a_e_colors_nolabels}) shows the Centaurs in the semimajor axis vs.~orbital eccentricity plane.  The observed Centaurs from Table (\ref{centaur_photometry}), and from Peixinho et al.~2003, 2012, Jewitt (2009) and Tegler, are  plotted with color-coding such that point-source red Centaurs (B-R $>$ 1.6) and blue Centaurs (B-R $\le$ 1.6) are distinguished by small red and blue circle symbols, respectively. Active Centaurs (from Table (\ref{centaur_photometry}) and Jewitt (2009), all of which have B-R $\le$ 1.6,  are shown as large yellow circles.  The orbital locations of all known JFCs (specifically, comets with 2 $\le T_J \le$ 3) and all known Centaurs, are marked as ocean green diamonds and grey circles, respectively.  The Figure shows that of the known active Centaurs all but one (167P) are found with perihelia between the orbital distances of Jupiter and Saturn, whereas inactive Centaurs  are observed to nearly the orbit of Neptune.   The KS test was used to compare the perihelion distance distribution of the active Centaurs with that of the entire known Centaur population.  By this test there is a 0.2\% likelihood of these distributions being consistent, corresponding to a 3.2$\sigma$ significant difference.  While there is an observational selection effect against the detection of coma in more distant objects, models of this selection effect by Jewitt (2009) suggest that it cannot account for the apparent concentration of the active Centaurs with perihelia $q \lesssim$ 10 AU.  As in Jewitt (2009), then, we  conclude that the active Centaurs tend to be those with smaller perihelion distances, as expected for activity driven by a thermal process.

The onset of activity near the 10 AU orbit of Saturn has two implications.  First,  these Centaurs presumably formed at larger distances, for otherwise their surfaces would have been devolatilized at formation and they would not reactivate upon returning to 10 AU.  This is probably consistent with most current solar system dynamical models (e.g. Nice model) in which strong radial mixing is an important feature; the active Centaurs set one small constraint on the degree of mixing.  

\begin{figure}
\epsscale{1.0}
\begin{center}
\plotone{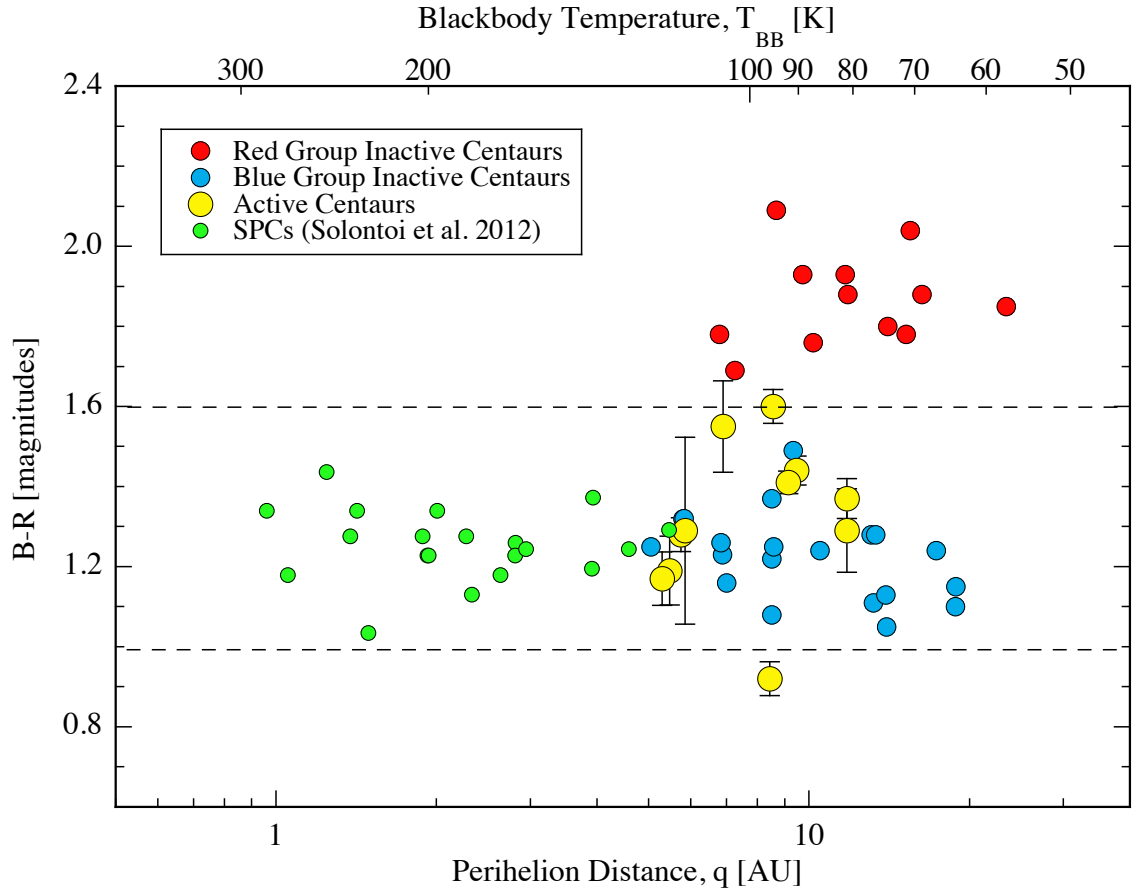}
\caption{B-R color index vs. perihelion distance for active Centaurs (yellow), red and blue inactive Centaurs and Jupiter family comets (green)  from Solontoi et al~(2012).  The upper horizontal axis shows the spherical blackbody  temperature.  The upper dashed horizontal line shows the location of the color gap between red and neutral groups as identified by Peixinho et al.~(2012) while the lower dashed line shows the B-R color of the Sun.  \label{BR_vs_q_Cen_Comets}
} 
\end{center} 
\end{figure}

Second, for activity to start as far out as $\sim$10 AU requires another volatile or a different process, since crystalline water ice sublimates negligibly at this distance.  (The upper axis of Figure (\ref{BR_vs_q_Cen_Comets}) shows, for reference, the spherical blackbody equilibrium temperature, calculated from $T_{BB}$ = 278 $q^{-1/2}$ K.  This is a lower limit to the surface temperature, while an upper limit is set by the subsolar temperature,  $T_{SS} = \sqrt 2 T_{BB}$.)  Supervolatile ices including CO$_2$ and CO could also drive mass loss at 10 AU if exposed to the heat of the Sun.  However, CO is so volatile that it would  sublimate strongly at 20 AU and even 30 AU distances, leaving no explanation for the $\sim$10 AU critical distance for the onset, as seen in Figure (\ref{a_e_colors_nolabels}).   Carbon dioxide is important in some cometary nuclei and must be present in the Centaurs.  Neither it nor CO is likely to be present as bulk ice but could be incorporated as clathrates or trapped in amorphous ice. The thermodynamic properties of clathrates are largely those of the host water molecule cage, so that activity at 10 AU would again be unexplained, leaving amorphous ice as the most plausible structure.  

Crystallization of amorphous ice, with the concommitant release of trapped volatiles, is the leading candidate process (Jewitt 2009). The latter obtained a simple criterion for the crystallization of amorphous ice exposed on the surface of an incoming Centaur. The resulting crystallization distances, 7 $\le r_h \le$ 14 AU, overlap the range of distances at which we observe activity in the Centaurs.  More sophisticated heat transport models reveal the effect of obliquity, $\psi$,  with complete crystallization out to $r_H \sim$10 AU for $\psi$ = 0\degr~and out to $r_H \sim$ 14 AU at $\psi$ = 90\degr~(Guilbert-Lepoutre 2012).  Therefore, even activity in the Centaur with the largest perihelion distance (167P/CINEOS at $q$ = 11.8 AU) is  consistent with a crystallization origin, provided the obliquity of this object is large.   

Figure (\ref{BR_vs_q_Cen_Comets}) shows the B-R colors of the Centaurs  and active JFCs as a function of the perihelion distance.  The active JFCs (green diamonds) have a mean B-R = 1.22$\pm$0.02 (N=26) indistinguishable from that of the active Centaurs (yellow circles; mean B-R = 1.30$\pm$0.05 (N=12)), within the uncertainties of measurement.  But the inactive Centaurs (blue circles) are on average much redder (mean B-R = 1.47$\pm$0.06 (N=32)) than either the active Centaurs or the JFCs.  As remarked above (c.f.~Figure \ref{BR_histo}, Peixinho et al.~2012), their colors appear bimodally distributed because of the presence of very red objects (B-R $\sim$ 1.9$\pm$0.1) that are not present in the other populations.  Taken together, Figures  (\ref{a_e_colors_nolabels}) and (\ref{BR_vs_q_Cen_Comets}) suggest that the transition from bimodal Centaur colors to unimodal colors begins at perihelion distances ($q \lesssim$ 10 AU) similar to the perihelion distances at which Centaur activity begins.  It is natural to suspect, then, that the disappearance of the ultrared matter is connected to  activity in the Centaurs.

\subsection{Blanketing and the Ultrared Matter}

A key observation from the present work is that the disappearance of ultrared matter  and the emergence of Centaur and JFC activity occur over  the same  range of perihelion distances, beginning at about 10 AU. This suggests that cometary activity itself, beginning in the Centaurs, is responsible for the disappearance of the ultrared matter.    Possible mechanisms include thermodynamic instability, ejection and burial (or ``blanketing'') of ultrared material (Jewitt 2002).   In support of this inference, numerical integrations of Centaur orbits show that ultrared objects have statistically spent less time inside $r_H <$ 9.5 AU than have neutral objects (Melita and Licandro 2012).

The low albedos and bland spectra of many middle and outer solar system objects have lead to a general expectation that organics, specifically irradiated organics, are optically important in these bodies (e.g.~Cooper et al.~2003). Indeed, recent measurements unambiguously reveal organic molecules on the surface of former Kuiper belt object 67P/Churyumov-Gerasimenko (Capaccionni et al.~2015, Goesmann et al.~2015, Wright et al.~2015).  The optical response of organic compounds to energetic bombardment is a strong function of composition.  Some, for example the simple organics methane (CH$_4$), methanol (CH$_3$OH) and benzene (C$_6$H$_6$) show increased reddening following irradiation by energetic particles (Brunetto et al.~2006).   Conversely, the complex, high molecular-weight hydrocarbons asphaltite and kerite (for which no simple chemical formulae can be given) become less red when irradiated, owing to the destruction of chemical bonds and the loss of hydrogen leading to carbonization of the material  (e.g.~Moroz et al.~2004).  They also show a simultaneous increase in the optical reflectivity (from $\sim$0.05 to $\sim$0.12) upon irradiation, interpreted as due to the graphitization of the material resulting from hydrogen loss.   The sense of these changes (towards less red materials having higher albedos) is opposite to the trend established in the Kuiper belt  (where redder objects have the higher albedos; Lacerda et al.~2014).  Interestingly, methanol has been spectroscopically reported in the ultrared (and inactive) Centaur 5145 Pholus (Cruikshank et al.~1998) and in the KBOs 2002 VE95 and 2004 TY364 (Merlin et al.~2012).  Brown et al.~(2011) have suggested that the ultrared matter could be caused by the irradiation of compounds including ammonia, which they believe might exist in certain regions of the Kuiper belt.  In any event, it is quite conceivable that the optical responses of different organics to energetic particle irradiation could be the cause of the wide dispersion of colors in the outer solar system (Figure \ref{color_field}), with ultrared surfaces from irradiated simple organics and more neutral objects  from irradiated complex, high molecular weight compounds.   

In this scenario, why would the ultrared surfaces disappear on objects approaching the Sun from the Kuiper belt?  Thermal instability (e.g.~sublimation) is one possibility.  However, the high molecular weight hydrocarbons are not volatile (Brunetto et al.~2006), particularly at 10 AU where the spherical blackbody temperature is only $T_{BB}$ = 88 K. It is difficult to see how they could sublimate away. Instead, we prefer an explanation in which outgassing activity destroys the ultrared matter, either by ejection or by blanketing of the nucleus by sub-orbital (``fallback'') debris (Jewitt 2002).  Since close-up observations show that the nuclei of comets are extensively shrouded in fine particulates, we focus on  blanketing  as the more likely mechanism for the change of color observed in Figures (\ref{color_field}) and (\ref{BR_vs_q_Cen_Comets}).  

Our assumption is that the ``primordial'' surfaces of KBOs consist of irradiated organics in a layer probably $\sim$1 m thick (Cooper et al.~2003), with  compositional differences (themselves products of different formation locations in the protoplanetary disk) responsible for the dispersion of colors from neutral to ultrared.  Once outgassing from sub-surface ice begins, this surface layer is obscured from view by blanketing with ``fresh'', un-irradiated material expelled from beneath.  Red and neutral irradiated organics alike are buried by fallback.

The time needed  to blanket a spherical nucleus of radius $r_n$ to depth $\Delta \ell$ with dust mass loss at total rate $dM/dt$ is

\begin{equation}
\tau_B = \frac{4\pi r_n^2 \rho \Delta \ell}{f_B (dM/dt)}
\label{taub}
\end{equation}

\noindent where $\rho$ is the density of the material and $f_B$ is the fraction of the ejecta, by mass, which falls back to the surface.  Jewitt (2002) used a variant of Equation (\ref{taub}) to estimate the timescale for formation of an insulating refractory mantle capable of suppressing sublimation of buried ice  (a so-called rubble mantle).  Such a mantle would need to be at least several centimeters thick, comparable to the diurnal thermal skin depth, to suppress sublimation.  However, a much thinner  deposited layer is sufficient to hide the underlying material from view and the timescales we derive here are thus very short compared to those needed to build an insulating rubble mantle.  We take $\Delta \ell$ = 10 $\mu$m  (corresponding to about 20 times the wavelength of observation), appropriate for an opaque, organic-rich material (Brunetto and Roush 2008) as found on the surface of the nucleus of comet 67P/Churyumov-Gerasimenko (Capaccioni et al.~2015).  We calculated values of $f_B$ according to the prescription in Jewitt (2002) for dust ejection through the action of gas drag (for which the terminal speed is related to the inverse square root of the particle size). At $r_H$ = 8 AU, we obtain $f_B = 4\times10^{-4}~r_n^{1.1}$, with $r_n$ expressed in kilometers.  Larger nuclei have larger capture fractions, all else being equal, because they have larger gravitational escape speeds.    Substituting for $f_B$ into Equation (\ref{taub}) gives

\begin{equation}
\tau_B~(yr) = 10~r_n^{0.9} \left[\frac{dM}{dt}\right]^{-1}
\label{taub2}
\end{equation}

\noindent with $r_n$ expressed in kilometers and $dM/dt$ in kg s$^{-1}$.  

Values of $r_n$ and  $dM/dt$ have been reported for several Centaurs based on the interpretation of integrated-light photometry.  For example, the large Centaur 29P/Schwassmann-Wachmann 1 has radius $r_n$ = 30$\pm$3 km (Schambeau et al.~2015) and a range of recent dust production rate estimates from 430 kg s$^{-1}$ to 1170 kg s$^{-1}$ (Fulle 1992, Ivanova et al.~2011, Shi et al.~2014).   Centaur P/2011 S1 has $r_n \le$ 3.9 km, $dM/dt$ = 40 to 150 kg s$^{-1}$ (Lin et al.~2014).  For P/2004 A1 the corresponding numbers are $r_n \le$ 3.5 km, $dM/dt \sim$ 130 kg s$^{-1}$ (Epifani et al.~2011), while for P/2010 C1 they are $r_n \le$ 4.8 km, $dM/dt$ = 0.1 to 15 kg s$^{-1}$ (Epifani et al.~2014).   

\begin{figure}
\epsscale{0.95}
\begin{center}
\plotone{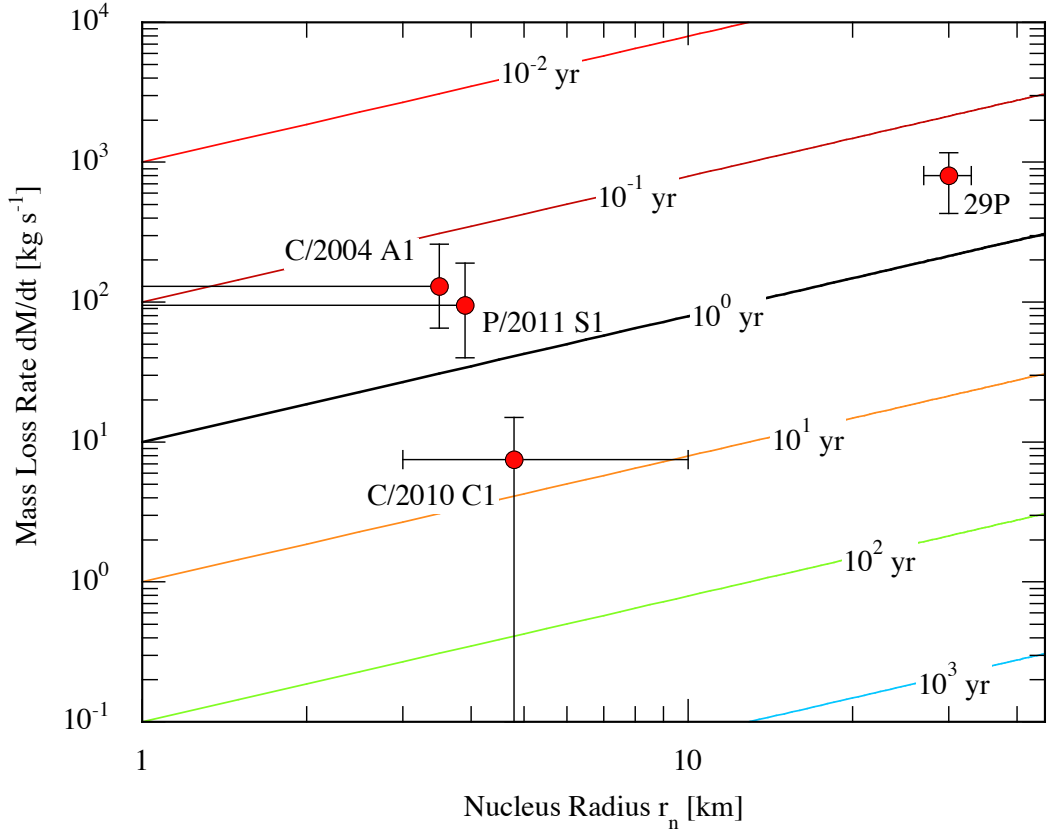}
\caption{Blanketing timescale  as a function of the nucleus radius in kilometers and the dust mass loss rate in kg s$^{-1}$.  Lines show the timescales in years (from Equation \ref{taub2}), as marked.  Circles denote measurements of four objects with their error bars.  Bars that extend to the left hand axis indicate that the radius measurements are reported upper limits. \label{dmbdt_vs_rn}
} 
\end{center} 
\end{figure}

The blanketing  timescales computed using these numbers with Equation (\ref{taub2}) are shown in Figure (\ref{dmbdt_vs_rn}).  Clearly, the  timescales are very uncertain, given the difficulties inherent in estimating $dM/dt$ from photometry and in calculating $f_B$ when neither the nucleus shape, rotation nor ejection mechanism can be well-specified.  In addition, Equations (\ref{taub}) and (\ref{taub2}) almost certainly underestimate $\tau_B$, perhaps by an order of magnitude or more, because Centaur activity is typically episodic and because fallback ejecta will not, in general, be uniformly distributed across the Centaur surface, with some areas taking longer to blanket than others.  Nevertheless, even with these caveats in mind, it is evident from Figure (\ref{dmbdt_vs_rn}) that  $\tau_B$ is very short compared to the 10$^6$ to 10$^7$ yr dynamical lifetimes of the Centaurs. As in Jewitt (2002), we conclude that blanketing of the surface by suborbital debris is an inevitable  and fast-acting consequence of outgassing activity.  

The demise of the ultrared matter, like the rise of Centaur activity, does not occur sharply at $q = $10 AU but is spread over a range of perihelion distances inwards to $\sim$7 AU (Figure \ref{BR_vs_q_Cen_Comets}).  This is consistent with simulations by Guilbert-Lepoutre (2012) which show that crystallization of amorphous ice in Centaurs can occur over a range of distances from $\sim$14 AU to $\sim$7 AU. This is because the surface temperature is influenced by the magnitude and direction of the spin vector (as well as by the body shape, which was not modeled). Activity and blanketing could also be delayed if the thermal diffusivity of the material or the thickness of the irradiated layer vary from object to object, as seems likely.     However, soon after activation, any object should promptly lose its original surface in this way.  Fallback blanketing  can account for the absence of ultrared matter in the comae and on the nuclei of active short- and long-period comets, where the observed material has been excavated from beneath the meter-thick  radiation damaged surface layer.   The Jovian Trojans, while they are not now active, exist close enough to the Sun that any exposed ice is unstable on long-timescales.  At $r_H$ = 5 AU ice should be depleted down to depths of meters or more (Guilbert-Lepoutre 2014) and, if the Trojans were once briefly closer to the Sun (Nesvorny et al.~2013), to even greater depths.   In the fallback blanketing scenario, the presence of ultrared matter indicates the absence of past or current activity.  However, we cannot conclude the opposite, namely that  neutral-group objects have necessarily been active leading to the burial of their irradiated surfaces.  Neutral group objects  in the Kuiper belt are very unlikely to have experienced  activity because their equilibrium temperatures are  low ($\sim$40 K).  Instead, the existence of a wide dispersion of surface colors in the Kuiper belt presumably reflects real compositional differences between bodies.  

An early model similar to this one invoked a competition between cosmic ray reddening and impact resurfacing to explain the color diversity in the Kuiper belt (Luu and Jewitt 1996).  The ``resurfacing model'' produced dramatic color variations only when the timescales for optical reddening and impact resurfacing were comparable, otherwise the color equilibrium would settle to one extreme or the other.  This model was later observationally rejected (Jewitt and Luu 2001), both because the measured color dispersion is larger than the model can produce and because expected azimuthal color variations on the KBOs were not detected.  The impact resurfacing model further struggles to account for Kuiper belt properties discovered since its formulation, notably the existence of the color-distinct cold Classical KBO (low inclination) population.  In the present context, the timescale for global impact resurfacing is assumed to be long compared to the timescale for reddening and the color diversity in the Kuiper belt has a compositional, not impact-caused, origin.  As the perihelion of an escaped KBO diffuses inward to the Sun, the near-surface temperatures rise until they are sufficient to trigger outgassing, whereupon blanketing of the surface by sub-orbital fallback debris is nearly immediate.

A few simple tests and consequences of the fallback blanketing hypothesis can be imagined.  First, the existence of ultrared comets or active Centaurs  would challenge the hypothesis directly, given the short timescales for blanketing indicated by Equation (\ref{taub2}).  Future observations should systematically search for  objects which are both outgassing and ultrared.  Second,  steep surfaces and outcrops on cometary nuclei might resist the accumulation of suborbital debris relative to more nearly horizontal surfaces.  High resolution color images could then reveal ultrared matter surviving on cliffs and outcrops  (e.g.~in 67P/Churyumov-Gerasimenko data from the ESA Rosetta spacecraft) and should be sought.  Third, on ultrared Kuiper belt objects and Centaurs, fresh impact craters  that are deeper than a few meters (i.e.~corresponding to crater diameters larger than perhaps 10 or 20 m)  should possess dark, neutral rims and rays consisting of  matter excavated from beneath the irradiated layer.  Images of the post-Pluto Kuiper belt objects to be visited by NASA's New Horizons spacecraft might be able to test this possibility, provided they have surfaces which are ultrared.

Grundy (2009) noted that ultrared colors could be produced  by fine tuning the wavelength dependence of the optical depths of ice particles through the addition of selected organic absorbers.  Sublimation of the ice could then cause the disappearance of the coloration.  Indeed, at 8 AU the thermal equilibrium sublimation mass flux of a flat water ice surface oriented perpendicular to sunlight is $F_s \sim$ 2$\times$10$^{-10}$ kg m$^{-2}$ s$^{-1}$, corresponding to a surface recession rate $F_s/\rho \sim$ 2$\times$10$^{-13}$ m s$^{-1}$, with ice density $\rho$ = 10$^3$ kg m$^{-3}$.  A 1 $\mu$m ice grain would take $\sim$5$\times$10$^6$ s (1 month) to sublimate away,   consistent with the prompt removal of ultrared matter on an object at this distance.  While we cannot rule it out, this  mechanism does rely upon arbitrary assumptions about the optical properties of the absorbing materials and about the sizes of the ice grains that are needed to guarantee strong wavelength-dependent optical depth effects. Moreover, we note that terrestrial frosts and snows  darken but do not become strongly colored when contaminated, reflecting the difficulty inherent in fine tuning the wavelength-dependent optical depth.  Neither have ultrared colors been noted on the icy surfaces of the outer planet satellites.  

\clearpage

\section{Summary}

We examined the optical colors of   short-lived, small-body populations originating in the Kuiper belt and Oort cloud comet reservoirs.  \begin{enumerate}

\item Ultrared matter, abundant in the Kuiper belt and Centaur populations, is less common (at the $\sim$95\% level of confidence) in active Centaurs. It is not detected in the active short-period or  long-period comets, in either their comae or nuclei. 

\item The onset of activity in the Centaurs  and the depletion of ultrared matter from the Centaur population both begin at perihelion distances $q \sim$ 10 AU.  This coincidence in distance suggests a connection between the two, namely that cometary activity  is itself responsible for the disappearance of ultrared matter. 

\item A plausible mechanism is the blanketing of ultrared surface material by an optically thick layer of fallback ejecta excavated from beneath the irradiated surface crust. Blanketing is a natural and probably unavoidable consequence of cometary activity, occurring on  timescales ($\sim$0.1 to 10 yr, for the cases considered here) that are very short compared to the dynamical lifetimes of the Centaurs and Jupiter family comets.

\item We find no significant difference between the mean optical colors of the dust in short-period (Kuiper belt) and long-period (Oort cloud) comets, or between the colors of the dust and the underlying nuclei in the comets of either group.  Neither do we find any correlation between the optical colors and the heliocentric or perihelion distances of the comets.  The latter shows that the  weighted mean particle size is always optically large (i.e.~scattering cross-sections are geometric), regardless of distance from the Sun and that ice grains in distant comets cannot be detected by their influence on the optical color.

\end{enumerate}

\acknowledgments
Masateru Ishiguro, Raquel Nuno, Hilke Schlichting, Chris Snead and Sid Grollix helped with the observations. Jing Li, Yoonyoung Kim and the anonymous referee made helpful comments on the manuscript.  We thank Luca Ricci (LRIS) and Julie Renaud-Kim (Keck) for assistance. This work was supported by a grant to DCJ from NASA's Solar System Observations program.

\clearpage

\begin{deluxetable}{lllllcc}
\tablecaption{Long Period Comet Orbital Elements 
\label{high_q_orbits}}
\tablewidth{0pt}
\tablehead{
\colhead{N}    & \colhead{Object}    & \colhead{$q$\tablenotemark{a}} & \colhead{$a$\tablenotemark{b}} & \colhead{$e$ \tablenotemark{c}}  
& \colhead{$i$\tablenotemark{d}} & \colhead{$T_P$\tablenotemark{e}} }
\startdata
1 & C/2003 A2 (Gleason) & 11.427 & -1643.9 & 1.007 & 8.1 & 2003 Nov 04.0 \\
2 & C/2004 D1 (NEAT)     & 4.975 & -3056.1 & 1.002 & 45.5 & 2006 Feb 10/8 \\
3 & C/2006 S3 (Loneos) & 5.131 & -1672 & 1.0031 & 166.0 & 2012 Apr 16.5 \\
4 & C/2007 D1 (LINEAR) & 8.793 & -30938 & 1.000 & 41.5 & 2007 Jun 19.0 \\
5 & C/2008 S3 (Boattini) & 8.019 & -7505.6 & 1.001 & 162.7 & 2011 Jun 05.9 \\
6 & C/2009 T1 (McNaught) & 6.220 & 3837.1 & 0.998 &89.9 & 2009 Oct 08.2 \\
7 & C/2010 D4 (WISE) & 7.148 & 64.656 & 0.889 & 105.7 & 2009 Mar 30.9 \\
8 & C/2010 DG56 (WISE)   		&  1.591 & 67.525 & 0.976 & 160.4 & 2010 May 15.6 \\
9 & C/2010 L3 (Catalina) & 9.882 & 12180.5 & 0.999 & 102.6 & 2010 Nov 10.4 \\
10 & C/2010 U3 (Boattini) & 8.469 & -7071.7 & 1.001 & 55.42 & 2019 Feb 26.3 \\
11 & C/2011 Q1 (PANSTARRS) & 6.780 & 3285.0 & 0.998 & 94.9 & 2011 Jun 29.4 \\
12 & C/2012 A1 (PANSTARRS) & 7.605 & -7505.85 & 1.001 & 120.9 & 2013 Nov 29.3 \\
13 & C/2012 E1 (Hill) &7.503 & 3761.43 & 0.998 & 122.5 & 2011 Jul 04.0 \\
14 & C/2012 K8 (Lemmon) & 6.464 & -2021.4 & 1.003 & 106.1 & 2014 Aug 19.2 \\
15 & C/2012 LP26 (Palomar) & 6.534 & 3954.3 & 0.998 & 25.4 & 2015 Aug 17.5 \\
16 & 2013 AZ60 & 7.911 & 991.67 & 0.992 & 16.5 & 2014 Nov 22.2 \\
17 & C/2013 E1 (McNaught) & 7.782 & -3133.88 & 1.003 & 158.7 & 2013 Jun 12.1 \\
18 & C/2013 H2 (Boattini) & 7.499 & -2657.52 & 1.003 & 128.4 & 2014 Jan 23.2 \\
19 & 2013 LD16            &   2.545 & 80.008 & 0.968 & 154.7 & 2013 Oct 14.8  \\

20 & C/2013 P3 (Palomar) & 8.646 & 9.99e99 & 1. & 93.9 & 2014 Nov 24.1 \\
21 & C/2014 AA52 (CATALINA) 	&  2.003 & -5503.2 & 1.000 & 105.2 & 2015 Feb 27.6 \\
22 & C/2014 B1 (Schwarz) & 9.531 & -642.7 & 1.015 & 28.4 & 2017 Sep 06.3 \\
23 & C/2014 R1 (Borisov) & 1.345 & 181.9 & 0.993 & 9.9 & 2014 Nov 19.2 \\
24 & C/2014 W6 (Catalina) & 3.088 & -1659 & 1.0019 & 53.6 & 2015 Mar 19.0 \\
25 & C/2014 XB8 (PANSTARRS) & 3.011 & 1902749 & 0.999998 & 149.8 & 2015 Apr 05.5 \\
26 & C/2015 B1 (PANSTARRS & 3.700 & 9.99e99 & 1. & 20.8 & 2016 Sep 20.9 \\


\enddata


\tablenotetext{a}{ Perihelion distance, AU}
\tablenotetext{b}{ Orbital semimajor axis, AU}
\tablenotetext{c}{ Orbital eccentricity}
\tablenotetext{d}{ Orbital inclination, degree}
\tablenotetext{e}{ Date of perihelion}


\end{deluxetable}

\clearpage

\begin{deluxetable}{lllclccccc}
\tablecaption{Long Period Comet Observational Geometry 
\label{comet_geometry}}
\tablewidth{0pt}
\tablehead{
\colhead{N} & \colhead{Object}  & \colhead{UT Date}   & \colhead{$r_H$\tablenotemark{a}} & \colhead{$\Delta$\tablenotemark{b}}  
& \colhead{$\alpha$\tablenotemark{c}} }
\startdata
1 & C/2003 A2 (Gleason)					& 2005 Jan 15  &  11.640 & 10.921 & 3.4 \\
2 & C/2004 D1 (NEAT)            & 2005 Jan 15 &   5.799 & 5.041 & 9.0 \\
3 & C/2006 S3 (Loneos) & 2015 Feb 18 & 9.011 & 8.164 & 3.4 \\
4 & C/2007 D1 (LINEAR) 		& 2010 Mar 17  & 10.490 & 9.587 & 2.4 \\
5 & C/2008 S3 (Boattini)                                	& 2010 Aug 10   & 8.222 & 8.055 & 7.0 \\
  & ~~~~~~~~~~''                                      		& 2010 Sep 10 &   8.182 & 7.505 & 5.5  \\
6 & C/2009 T1 (McNaught) & 2011 Jan 30 &   7.024 & 6.894 & 8.0 \\
7 & C/2010 D4 (WISE)                                  & 2010 Sep 10 &   7.822 & 8.114 & 6.9 \\
8 & C/2010 DG56 (WISE) & 2010 Sep 10 & 2.210 & 1.226 & 7.3 \\
9 & C/2010 L3 (Catalina) 				& 2010 Sep 10  & 9.889 & 10.094 & 5.7\\
10 & C/2010 U3 (Boattini) 	& 2011 Jan 30 &   17.982 & 18.003 & 3.1  \\
  & ~~~~~~~~~~''                                       	 	& 2012 Oct 13 &   15.382 & 14.487& 1.7  \\
  & ~~~~~~~~~~''                                       		& 2012 Oct 14 &   15.378 & 14.487 & 1.6  \\
11 & C/2011 Q1 (PANSTARRS) & 2012 Oct 13 &   7.449 & 6.961 & 6.9  \\
  & ~~~~~~~~~~''                                             & 2012 Oct 14 &   7.452 & 6.998 & 7.0  \\
12 & C/2012 A1 (PANSTARRS)          & 2014 Feb 26 &   7.621 & 7.333 & 7.3 \\
13 & C/2012 E1 (Hill) & 2014 Feb 26 &    9.567 & 9.048 & 5.2 \\
14 & C/2012 K8 (Lemmon) & 2012 Oct 13 &   7.872 & 7.695 & 7.2 \\
15 & C/2012 LP26 (Palomar) & 2012 Oct 13 & 9.378  & 10.174 & 3.5 \\
  & ~~~~~~~~~~''                				& 2012 Oct 14 &  9.374  & 10.176 & 3.5 \\
  & ~~~~~~~~~~''   					& 2014 Feb 27 &   7.442 & 7.677 & 7.3 \\
16 & 2013 AZ60 & 2014 Feb 26 &   8.077 & 7.214 & 3.6 \\
17 & C/2013 E1 (McNaught) & 2014 Feb 26 &    7.944 & 6.983 & 1.8 \\

18 & C/2013 H2 (Boattini)                                & 2014 Feb 27 &    7.502 & 7.384 & 7.6 \\
19 & 2013 LD16\tablenotemark{d} & 2014 Feb 27 & 2.915 & 2.001 & 9.1 \\
20 & C/2013 P3 (Palomar) & 2013 Oct 01 &   8.984 & 8.101 & 3.2 \\
21 & C/2014 AA52 (CATALINA) & 2014 Feb 26 &  4.483 & 3.563 & 5.2 \\
22 & C/2014 B1 (Schwarz)          & 2014 Feb 26 &   11.875 & 11.462 & 4.4 \\

23 & C/2014 R1 (Borisov) & 2015 Feb 18 & 1.875 & 1.882 & 30.5  \\
24 & C/2014 W6 (Catalina) & 2015 Feb 18 & 3.100  & 2.282 & 12.0  \\
25 & C/2014 XB8 (PANSTARRS) & 2015 Feb 17 & 3.047 & 3.239 & 17.8 \\
26 & C/2015 B1 (PANSTARRS) & 2015 Feb 18 & 6.145 & 5.203 & 3.0   \\

\enddata


\tablenotetext{a}{Heliocentric distance, AU}
\tablenotetext{b}{Geocentric distance, AU}
\tablenotetext{c}{Phase angle, degree}

\end{deluxetable}

\clearpage

\begin{landscape}
\begin{deluxetable}{lllcccccccll}
\tabletypesize{\scriptsize}
\tablecaption{Long Period Comet Photometry
\label{lpc_photometry}}
\tablewidth{0pt}
\tablehead{
\colhead{N} & \colhead{Object} & \colhead{UT Date} & \colhead{Note}\tablenotemark{a} & \colhead{$\phi$}\tablenotemark{b}  & \colhead{$m_R$\tablenotemark{c}}  & \colhead{$B-V$} & \colhead{$V-R$} & \colhead{$R-I$} & \colhead{$B-R$} }
\startdata

1 & C/2003 A2 (Gleason)  & 2005 Jan 15 & E & 6.0 & 19.96$\pm$0.03 & 0.61$\pm$0.04 & 0.47$\pm$0.04 & 0.46$\pm$0.04 & 1.08$\pm$0.04 \\

2 & C/2004 D1 (NEAT) & 2005 Jan 15 & E &  6.0 & 17.54$\pm$0.03 & 0.82$\pm$0.05 & 0.43$\pm$0.04 & 0.51$\pm$0.04 & 1.25$\pm$0.05 \\
3 & C/2006 S3 (Loneos) & 2015 Feb 18 & E &  8.1 & 17.57$\pm$0.01  & 0.74$\pm$0.01 & 0.58$\pm$0.01 & -- & 1.32$\pm$0.02 \\
4 & C/2007 D1 (LINEAR) & 2010 Mar 17 & E &  5.4 & 18.85$\pm$0.02 & 0.75$\pm$0.02 & 0.44$\pm$0.02  & 0.41$\pm$0.03 & 1.19$\pm$0.03 \\
5 & C/2008 S3 (Boattini) & 2010 Aug 10 & E &  6.8 & 18.57$\pm$0.02 & 0.74$\pm$0.05 & 0.48$\pm$0.04  & 0.38$\pm$0.04 & 1.22$\pm$0.05\\
  & ~~~~~~~~~~''  & 2010 Sep 10 & E &  8.2  &  18.23$\pm$0.01 & 0.78$\pm$0.02 & 0.44$\pm$0.01  & 0.36$\pm$0.01& 1.22$\pm$0.02 \\
6 & C/2009 T1 (McNaught) & 2011 Jan 30 & E &  9.4 & 19.25$\pm$0.03 & 0.64$\pm$0.04 & 0.53$\pm$0.03  & 0.45$\pm$0.03 & 1.17$\pm$0.03\\
7 & C/2010 D4(WISE)  & 2010 Sep 10 & P &  8.0 & 20.91$\pm$0.05 & 0.74$\pm$0.07 & 0.46$\pm$0.05 & -- & 1.20$\pm$0.05 & -- \\
8 & C/2010 DG56 (WISE) & 2010 Sep 10 & E &  8.0 & 19.98$\pm$0.03 & 0.77$\pm$0.05 & 0.37$\pm$0.05 & -- & 1.14$\pm$0.05 \\
9 & C/2010 L3 (Catalina) & 2010 Sep 10 & E &  5.6 & 19.94$\pm$0.02 & 0.75$\pm$0.03 & 0.42$\pm$0.03  & 0.41$\pm$0.03 & 1.17$\pm$0.03 \\
10 & C/2010 U3 (Boattini) & 2011 Jan 30 & E &  9.4 & 20.09$\pm$0.05 & 0.82$\pm$0.05 & 0.48$\pm$0.04  & 0.32$\pm$0.05 & 1.30$\pm$0.06 \\
  & ~~~~~~~~~~''    & 2012 Oct 13 & E &  6.8 & 20.04$\pm$0.02 & 0.81$\pm$0.03 & 0.54$\pm$0.03 & -- & 1.35$\pm$0.03 \\
  & ~~~~~~~~~~''   & 2012 Oct 14 & E &  6.8 & 19.86$\pm$0.02 & 0.72$\pm$0.03 & 0.53$\pm$0.03 & -- & 1.25$\pm$0.03 \\
11 & C/2011 Q1 (PANSTARRS) & 2012 Oct 13 & E &  6.8 & 20.98$\pm$0.06 & 0.83$\pm$0.05 & 0.51$\pm$0.07 & -- & 1.34$\pm$0.07 \\
 & ~~~~~~~~~~''   & 2012 Oct 14 & E &  6.8 & 21.02$\pm$0.02 & 0.81$\pm$0.04 & 0.48$\pm$0.03 & -- & 1.29$\pm$0.04 \\
12 & C/2012 A1 (PANSTARRS) & 2014 Feb 26 & E &  8.0 & 18.84$\pm$0.02 & 0.74$\pm$0.01 & 0.45$\pm$0.02 & -- & 1.19$\pm$0.03 \\
13 & C/2012 E1 (Hill) & 2014 Feb 26 & E &  8.1 & 21.84$\pm$0.05 & 0.79$\pm$0.09 & 0.40$\pm$0.07 & -- & 1.19$\pm$0.09 \\
14 & C/2012 K8 (Lemmon) & 2012 Oct 13 & P &  6.8 & -- & 0.72$\pm$0.01 & -- & -- & -- \\
15 & C/2012 LP26 (Palomar)  & 2012 Oct 13 & E &  6.8 & 19.60$\pm$0.03 & 0.88$\pm$0.03 & 0.52$\pm$0.04 & -- & 1.40$\pm$0.04 \\
  & ~~~~~~~~~~''   & 2012 Oct 14 & E &  6.8 & 19.57$\pm$0.01 & 0.91$\pm$0.02 & 0.58$\pm$0.01 & -- & 1.49$\pm$0.01 \\
  & ~~~~~~~~~~''   & 2014 Feb 27 & E &  8.1& 19.12$\pm$0.01 & 0.78$\pm$0.02 & 0.45$\pm$0.01 & -- & 1.23$\pm$0.02 \\
16 & 2013 AZ60    & 2014 Feb 26 & E &  8.1 & 18.85$\pm$0.01 & 0.82$\pm$0.01 & 0.54$\pm$0.01 & -- & 1.36$\pm$0.01 \\
17 & C/2013 E1 (McNaught) & 2014 Feb 26 & E &  8.1 & 19.16$\pm$0.02 & 0.75$\pm$0.03 & 0.48$\pm$0.03 & -- & 1.23$\pm$0.03 \\
18 & C/2013 H2 (Boattini) & 2014 Feb 27 & E &  8.1 & 18.80$\pm$0.01 & 0.77$\pm$0.02 & 0.49$\pm$0.01 & -- & 1.26$\pm$0.02 \\
19 & 2013 LD16 & 2014 Feb 27 & E &  8.0 & 20.15$\pm$0.02 & 0.86$\pm$0.05 & 0.44$\pm$0.03 & -- & 1.30$\pm$0.02 \\
20 & C/2013 P3 (Palomar) & 2013 Oct 01 & E &  8.1 & 19.78$\pm$0.04 & 0.92$\pm$0.05 & 0.45$\pm$0.05 & -- & 1.37$\pm$0.06 & \\
21 & C/2014 AA52 (CATALINA) & 2014 Feb 26 & E &  8.0 & 18.36$\pm$0.02 & 0.77$\pm$0.02 & 0.41$\pm$0.02 & -- & 1.18$\pm$0.03 \\\\
22 & C/2014 B1 (Schwarz) & 2014 Feb 26 & E &  8.1& 19.20$\pm$0.02 & 0.85$\pm$0.03 & 0.58$\pm$0.03 & -- & 1.43$\pm$0.03 \\
23 & C/2014 R1 (Borisov) & 2015 Feb 18 & E &  8.1 & 15.40$\pm$0.01 & 0.81$\pm$0.01 & 0.46$\pm$0.01 & -- &1.27$\pm$0.01  \\
24 & C/2014 W6 (Catalina) & 2015 Feb 18 & E &  8.1  & 18.09$\pm$0.01 & 0.81$\pm$0.01 & 0.45$\pm$0.01 & -- & 1.26$\pm$0.01  \\
25 & C/2014 XB8 (PANSTARRS) & 2015 Feb 17 & E &  5.4 & 20.92$\pm$0.02 & 0.79$\pm$0.03 & 0.44$\pm$0.03 & -- & 1.23$\pm$0.03 \\
26 & C/2015 B1 (PANSTARRS) & 2015 Feb 18 & E &  8.1 & 19.51$\pm$0.01 & 0.78$\pm$0.01 & 0.45$\pm$0.01 & -- & 1.23$\pm$0.01    \\


 & Sun\tablenotemark{d} & -- & -- &  -- & -- & 0.64$\pm$0.02 & 0.35$\pm$0.01 & 0.33$\pm$0.01 & 0.99$\pm$0.02 \\

\enddata


\tablenotetext{a}{Morphology; P = point source, E = extended}
\tablenotetext{b}{Angular diameter of photometry aperture in arcsecond.} 
\tablenotetext{c}{Apparent red magnitude inside aperture of diameter $\phi$}
\tablenotetext{d}{Holmberg et al.~2006}
\end{deluxetable}
\end{landscape}
\clearpage

\begin{deluxetable}{lllllcc}
\tablecaption{Damocloid Orbital Elements 
\label{damo_orbits}}
\tablewidth{0pt}
\tablehead{
\colhead{N} & \colhead{Object}    & \colhead{$q$\tablenotemark{a}} & \colhead{$a$\tablenotemark{b}} & \colhead{$e$ \tablenotemark{c}}  
& \colhead{$i$\tablenotemark{d}} & \colhead{$T_P$\tablenotemark{e}} }
\startdata
1 & 2010 BK118 & 6.106 & 452.642 & 0.987 & 143.9 & 2012 Apr 30.2 \\
2 & 2010 OM101  & 2.129 & 26.403 & 0.919 & 118.7 & 2010 Oct 01.4 \\
3 & 2010 OR1 & 2.052 & 27.272 & 0.925 & 143.9 & 2010 Jul 12.5 \\
4 & 2012 YO6 & 3.304 & 6.339 & 0.479 & 106.9 & 2012 Jul 30.3 \\
5 & 2013 NS11 & 2.700 & 12.646 & 0.786 & 130.4 & 2014 Sep 26.2 \\
6 & 2013 YG48 & 2.024 & 8.189 & 0.753 & 61.3 & 2014 Mar 11.6 \\
7 & 2014 CW14 & 4.235 & 19.810 & 0.786 & 170.6 & 2014 Dec 23.4 \\
8 & 330759 (2008 SO218) & 3.546 & 8.148 & 0.565 & 170.4 & 2009 Dec 31.2 \\
9 & 336756 (2010 NV1) & 9.417 & 292.030 & 0.968 & 140.8 & 2010 Dec 13.2 \\
10 & 342842 (2008 YB3) & 6.487 & 11.651 & 0.443 & 105.0 & 2011 Mar 01.7 \\
11 & 418993 (2009 MS9) & 11.004 & 387.251 & 0.972 & 68.0 & 2013 Feb 11.7 \\

\enddata


\tablenotetext{a}{ Perihelion distance, AU}
\tablenotetext{b}{ Orbital semimajor axis, AU}
\tablenotetext{c}{ Orbital eccentricity}
\tablenotetext{d}{ Orbital inclination, degree}
\tablenotetext{e}{ Date of perihelion}


\end{deluxetable}


\clearpage

\begin{deluxetable}{lllrrr}
\tablecaption{Damocloid Observational Geometry 
\label{damo_geometry}}
\tablewidth{0pt}
\tablehead{
\colhead{N} & \colhead{Object}  & \colhead{UT Date}    & \colhead{$r_H$\tablenotemark{a}} & \colhead{$\Delta$\tablenotemark{b}}  
& \colhead{$\alpha$\tablenotemark{c}} }
\startdata
 
1 & 2010 BK118           & 2011 Jan 30 &  6.855 & 7.079 & 7.9 \\
   &  ~~~~~~~~~~''        & 2010 Sep 10 & 7.330 & 6.825 & 7.1 \\
2 & 2010 OM101         & 2010 Aug 10 &  2.208 & 1.780 & 26.8 \\
 &  ~~~~~~~~~~''                                 & 2010 Sep 10 &  2.143 & 1.469 & 24.4 \\
3 & 2010 OR1               & 2010 Aug 10 &  2.079 & 1.278 & 22.0 \\
 & ~~~~~~~~~~''                                   & 2010 Sep 10 &  2.164 & 1.254 & 15.1 \\
4 & 2012 YO6               & 2013 Oct 01 &  4.243 & 3.938 & 13.4 \\
5 & 2013 NS11            & 2014 Feb 26 &   3.312 & 2.813 & 16.1 \\
6 & 2013 YG48 &    2014 Feb 26 &  2.028 & 1.268 & 22.8 \\
 & ~~~~~~~~~~''                            &    2014 Feb 27 &  2.028 & 1.278 & 23.2 \\
7 & 2014 CW14 & 2014 Feb 26 &  4.770 & 3.845 & 4.7 \\
8 & 330759 (2008 SO218)           & 2010 Sep 10 &  3.937 & 3.197 & 11.0 \\    
9 & 336756 (2010 NV1)               & 2010 Aug 10 &  9.441 & 8.851 & 5.2 \\         
10 & 342842 (2008 YB3)                & 2014 Feb 26 &   7.998 & 7.609 & 6.7 \\
11 & 418993 (2009 MS9)                & 2010 Sep 10 &  11.886 & 11.553 & 4.6 \\

\enddata


\tablenotetext{a}{Heliocentric distance, AU}
\tablenotetext{b}{Geocentric distance, AU}
\tablenotetext{c}{Phase angle, degree}

\end{deluxetable}


\clearpage

\begin{landscape}
\begin{deluxetable}{cllcccccccll}
\tabletypesize{\scriptsize}
\tablecaption{Damocloid Photometry
\label{damo_photometry}}
\tablewidth{0pt}
\tablehead{
\colhead{N} & \colhead{Object} & \colhead{UT Date} &  \colhead{Note\tablenotemark{a}} & \colhead{$\phi$\tablenotemark{b}}  & \colhead{$m_R$\tablenotemark{c}}  & \colhead{$B-V$} & \colhead{$V-R$} & \colhead{$R-I$} & \colhead{$B-R$} }
\startdata
1 & 2010 BK118 & 2011 Jan 30 & P & 5.4 & 18.97$\pm$0.02 & 0.76$\pm$0.03 & 0.57$\pm$0.03  & 0.51$\pm$0.03 & 1.33$\pm$0.03\\
 & ~~~~~~~~~~''  & 2010 Sep 10 & P &  8.0 & 20.06$\pm$0.02 & 0.80$\pm$0.02 & 0.52$\pm$0.02  & 0.45$\pm$0.03 & 1.32$\pm$0.03\\
2 &  2010 OM101  & 2010 Aug 10 & P &  5.6 & 20.64$\pm$0.03 & 0.81$\pm$0.04 & 0.53$\pm$0.04 & -- & 1.34$\pm$0.05 & -- \\
 & ~~~~~~~~~~''   & 2010 Sep 10 & P &  5.6 & 19.94$\pm$0.07 & 0.77$\pm$0.07 & 0.69$\pm$0.07  & 0.36$\pm$0.07 & 1.46$\pm$0.07\\
3 & 2010 OR1 & 2010 Aug 10 & P &  6.8 & 18.94$\pm$0.02 & 0.76$\pm$0.03 & 0.52$\pm$0.02  & 0.48$\pm$0.04 & 1.28$\pm$0.04\\
 & ~~~~~~~~~~''  & 2010 Sep 10 & P &  8.0 & 19.97$\pm$0.02 & 0.81$\pm$0.04 & 0.51$\pm$0.04  & 0.43$\pm$0.02 & 1.32$\pm$0.03\\
4 & 2012 Y06 & 2013 Oct 01 & P &  8.0 & 21.59$\pm$0.05 & -- & --  &  0.34$\pm$0.05 & 1.32$\pm$0.05 \\
5 & 2013 NS11 & 2014 Feb 26 & P &  8.0 & 19.12$\pm$0.03 & 0.74$\pm$0.02 & 0.56$\pm$0.04 & -- & 1.30$\pm$0.03 \\
6 & 2013 YG48 & 2014 Feb 26 & P &  8.0 & 20.13$\pm$0.02 & 0.77$\pm$0.02 & 0.55$\pm$0.02 & -- & 1.32$\pm$0.02 \\
 & ~~~~~~~~~~''  & 2014 Feb 27 & P &  5.4 & 19.94$\pm$0.01 & 0.83$\pm$0.02 & 0.48$\pm$0.02 & -- & 1.31$\pm$0.02 \\
7 & 2014 CW14 & 2014 Feb 26 & P &  8.0 & 20.97$\pm$0.02 & 0.87$\pm$0.09 & 0.51$\pm$0.06 & -- & 1.38$\pm$0.07 \\
8 &  330759 (2008 SO218) & 2010 Sep 10 & P &  5.6 & 18.81$\pm$0.01 & 0.89$\pm$0.03 & 0.55$\pm$0.02  & 0.50$\pm$0.04 & 1.44$\pm$0.02\\
9 & 336756 (2010 NV1) & 2010 Aug 10 & P &  8.0 & 20.31$\pm$0.02 & 0.79$\pm$0.03 & 0.53$\pm$0.03 & 0.39$\pm$0.02 & 1.32$\pm$0.03 \\
10 & 342842 (2008 YB3) & 2014 Feb 26 & P &  8.0 & 18.35$\pm$0.01 & 0.74$\pm$0.02 & 0.51$\pm$0.02 & -- & 1.25$\pm$0.02 \\
11 & 418993 (2009 MS9) & 2010 Sep 10 &P &  3.8 & 20.79$\pm$0.04 & 0.84$\pm$0.04 & 0.52$\pm$0.04 & -- & 1.36$\pm$0.04 & \\

\\
 & Sun\tablenotemark{d} & -- & -- &   -- & -- & 0.64$\pm$0.02 & 0.35$\pm$0.01 & 0.33$\pm$0.01 & 0.99$\pm$0.02 \\


\enddata

\tablenotetext{a}{Morphology; P = point source, E = extended}
\tablenotetext{b}{Angular diameter of photometry aperture in arcsecond}
\tablenotetext{c}{Apparent red magnitude  measured in aperture of diameter $\phi$}
\tablenotetext{d}{Holmberg et al.~2006}

\end{deluxetable}
\end{landscape}


\clearpage

\begin{deluxetable}{lllllcc}
\tablecaption{Centaur Orbital Elements 
\label{centaur_orbits}}
\tablewidth{0pt}
\tablehead{
\colhead{N} & \colhead{Object}    & \colhead{$q$\tablenotemark{a}} & \colhead{$a$\tablenotemark{b}} & \colhead{$e$ \tablenotemark{c}}  
& \colhead{$i$\tablenotemark{d}} & \colhead{$T_P$\tablenotemark{e}} }
\startdata
1 & 2001 XZ255                 & 15.476 & 16.026 & 0.034 & 2.6 & 1986 May 06.9 \\
2 & 2010 BL4                      & 8.573 & 18.534 & 0.537 & 20.8 & 2009 Jul 27.1 \\
3 & 2010 WG9                    & 18.763 & 53.729 & 0.651 &70.2 & 2006 Feb 20.4 \\
4 & 2013 BL76                   & 8.374 & 1216.0 & 0.993 & 98.6 & 2012 Oct 27.7 \\
5 & 2014 HY123 &                6.998 & 18.552 & 0.623 & 14.0 & 2017 Jan 08.1 \\
6 & 2015 CM3 &                  6.784 & 13.182 & 0.485 & 19.7 & 2013 Apr 26.0 \\
7 & 32532 Thereus (2001 PT13) & 8.513 & 10.615 & 0.198 & 20.4 & 1999 Feb 12.0 \\
8 & 148975 (2001 XA255)                & 9.338 & 29.617 & 0.685 & 12.6 & 2010 Jun 19.2 \\
9 & 160427 (2005 RL43)                        & 23.449 & 24.550 & 0.045 & 12.3 & 1991 Sep 13.6 \\
10 & 433873 (2015 BQ311) &              5.051 & 7.140 & 0.293 & 24.5 & 2007 Oct 16.1 \\
11 & C/2011 P2  (PANSTARRS)                   & 6.148 & 9.756 & 0.370 & 9.0 & 2010 Sep 13.5 \\
12 & P/2011 S1 (Gibbs)      & 6.897 & 8.655 & 0.203 & 2.7 & 2014 Aug 20.2 \\
13 & C/2012 Q1 (Kowalski) & 9.482 & 26.154 & 0.637 & 45.2 & 2012 Feb 09.4 \\                  
14 & C/2013 C2 (Tenagra) & 9.132 & 15.993 & 0.429 & 21.3 & 2015 Aug 28.6 \\
15 & C/2013 P4 (PANSTARRS)  & 5.967 & 14.779 & 0.596 & 4.3 & 2014 Aug 12.4 \\
16 & 166P/NEAT &                 8.564 & 13.883 & 0.383 & 15.4 & 2002 May 20.8 \\
17 &167P/CINEOS             & 11.784 & 16.141 & 0.270 & 19.1 & 2001 Apr 10.0 \\

\enddata


\tablenotetext{a}{ Perihelion distance, AU}
\tablenotetext{b}{ Orbital semimajor axis, AU}
\tablenotetext{c}{ Orbital eccentricity}
\tablenotetext{d}{ Orbital inclination, degree}
\tablenotetext{e}{ Epoch of perihelion}

\end{deluxetable}


\clearpage

\begin{deluxetable}{lllclccccc}
\tablecaption{Centaur Observational Geometry 
\label{centaur_geometry}}
\tablewidth{0pt}
\tablehead{
\colhead{N} & \colhead{Object}  & \colhead{UT Date}    & \colhead{$r_H$\tablenotemark{a}} & \colhead{$\Delta$\tablenotemark{b}}  
& \colhead{$\alpha$\tablenotemark{c}} }
\startdata
1 & 2001 XZ255 & 2003 Jan 09 &     16.079 & 15.096 & 0.1      \\ 
2 & 2010 BL4 & 2010 Mar 17 &   8.632 & 8.017 & 5.4          \\
3 & 2010 WG9 &  2011 Jan 29 &    19.621 & 19.161 & 2.6         \\
4 & 2013 BL76 & 2013 Oct 01 &  8.610 & 7.627 & 1.3            \\
5 & 2014 HY123 &                2015 Feb 17 & 7.818 & 6.838 & 1.0 \\
6 & 2015 CM3 &                  2015 Feb 17 & 7.417 & 6.540 & 3.7  \\
7 & 32532 Thereus (2001 PT13) & 2007 Feb 20 &   10.792 & 11.233 & 4.6            \\
 & ~~~~~~~~~~''                                    & 2015 Feb 17 & 12.703 & 12.087 & 3.6 \\
8 & 148975 (2001 XA255)   & 2003 Jan 09 &   14.965 & 13.982 & 0.1       \\
9 & 160427 (2005 RL43)     &  2014 Feb 26 &   24.142 & 24.544 & 2.1          \\
10 & 433873 (2015 BQ311)  &              2015 Feb 17 & 8.888 & 7.944 & 2.0  \\

11 & C/2011 P2 (PANSTARRS)                     & 2014 Oct 22 & 8.412 & 7.475 & 2.4 \\
12 & P/2011 S1 (Gibbs) &  2014 Feb 26 &   6.914 & 7.398 & 6.9          \\
13 &  C/2012 Q1 (Kowalski)    &  2012 Oct 13 &    9.546 & 8.837 & 4.4       \\
 & ~~~~~~~~~~''                                     &  2012 Oct 14 &       9.546 & 8.848 & 4.4      \\
 & ~~~~~~~~~~''     &  2013 Oct 01 &    9.849 & 8.903 & 2.0         \\

14 & C/2013 C2 (Tenagra)   & 2014 Feb 26 &   9.355 & 8.478 & 3.0          \\
15 & C/2013 P4 (PANSTARRS) & 2014 Oct 22 & 5.980 & 5.015 & 2.5 \\
16 & 166P/NEAT &                 2015 Feb 17 & 15.640 & 14.700 & 1.1  \\
17 & 167P/CINEOS             & 2010 Sep 10 &  14.415 & 13.481 & 1.6           \\
 & ~~~~~~~~~~''                                    & 2012 Oct 13 &  15.332 & 14.398 & 1.3           \\

\enddata


\tablenotetext{a}{Heliocentric distance, AU}
\tablenotetext{b}{Geocentric distance, AU}
\tablenotetext{c}{Phase angle, degree}

\end{deluxetable}

\clearpage

\begin{landscape}
\begin{deluxetable}{lllccccccll}
\tabletypesize{\scriptsize}
\tablecaption{Centaur Photometry
\label{centaur_photometry}}
\tablewidth{0pt}
\tablehead{
\colhead{N} & \colhead{Object} & \colhead{UT Date} & \colhead{Note\tablenotemark{a}} &\colhead{$\phi$\tablenotemark{b}}  & \colhead{$m_R$\tablenotemark{c}}  & \colhead{$B-V$} & \colhead{$V-R$} & \colhead{$R-I$} & \colhead{$B-R$} }
\startdata

1 & 2001 XZ255 & 2003 Jan 09 	& P & 5.4 & 22.43$\pm$0.03 			& 1.36$\pm$0.05 & 0.68$\pm$0.05 & --& 2.04$\pm$0.05 \\
2 & 2010 BL4 & 2010 Mar 17 	& P & 5.4 & 21.18$\pm$0.03 			& 0.86$\pm$0.05 & 0.39$\pm$0.03  & 0.47$\pm$0.03 & 1.25$\pm$0.03\\
3 & 2010 WG9 & 2011 Jan 29 	& P & 9.4 & 20.92$\pm$0.02 			& 0.73$\pm$0.05 & 0.37$\pm$0.04 & -- & 1.10$\pm$0.05 & -- \\
4 & 2013 BL76 & 2013 Oct 01 	& P & 8.0 & 19.78$\pm$0.04 			& 0.92$\pm$0.05 & 0.45$\pm$0.05 & -- & 1.37$\pm$0.06 & -- \\
5 &  2014 HY123 & 2015 Feb 17 	&P& 5.4  & 20.03$\pm$0.02 & 0.67$\pm$0.05 & 0.49$\pm$0.02 & -- & 1.16$\pm$0.06 \\
6 & 2015 CM3 &   2015 Feb 17 	& P& 8.2 & 20.99$\pm$0.04 & 1.21$\pm$0.06 & 0.57$\pm$0.05 & -- & 1.78$\pm$0.06 \\
7 & 32532 Thereus (2001 PT13) & 2007 Feb 20 	&P & 5.4 & 19.90$\pm$0.02 		& -- & 0.41$\pm$0.05 & 0.43$\pm$0.05 & -- \\
 & ~~~~~~~~~~''                                 	& 2015 Feb 17 &P & 5.4 & 20.00$\pm$0.02 & 0.77$\pm$0.03 & 0.45$\pm$0.03 & -- & 1.22$\pm$0.03 \\
8 & 148975 (2001 XA255)   & 2003 Jan 09 	& P & 5.4 & 22.35$\pm$0.03 			& 0.81$\pm$0.05 & 0.68$\pm$0.05 & 0.44$\pm$0.05 & 1.49$\pm$0.05 \\
9 & 160427 (2005 RL43)  & 2014 Feb 26 	& P & 8.0 &21.39$\pm$0.05 				& 1.12$\pm$0.04 & 0.73$\pm$0.06 & -- & 1.85$\pm$0.05 \\
10 & 433873 (2015 BQ311) &              2015 Feb 17 	& P & 5.4 & 21.43$\pm$0.02 & 0.85$\pm$0.03 & 0.40$\pm$0.03 & -- & 1.25$\pm$0.03 \\

11 & C/2011 P2 (PANSTARRS)  & 2014 Oct 22 	&  E &2.7 & 22.18$\pm$0.04 & 0.81$\pm$0.06 & 0.43$\pm$0.05 & -- & 1.24$\pm$0.07 \\
 & ~~~~~~~~~~''              &    &      E &              5.4 & 22.11$\pm$0.03 & 1.00$\pm$0.05 & 0.21$\pm$0.05 & -- & 1.21$\pm$0.05 \\
 & ~~~~~~~~~~''                   &     &     E &            8.2 & 22.04$\pm$0.05 & 0.93$\pm$0.06 & 0.33$\pm$0.05 & -- & 1.26$\pm$0.06 \\
12 & P/2011 S1 (Gibbs) & 2014 Feb 26 & E & 8.0 & 20.55$\pm$0.03 			& 0.96$\pm$0.11 & 0.59$\pm$0.03 & -- & 1.55$\pm$0.11 \\

13 & C/2012 Q1 (Kowalski)  & 2012 Oct 13 & E & 13.6 & 19.60$\pm$0.03 				& 0.88$\pm$0.03 & 0.52$\pm$0.04 & -- & 1.40$\pm$0.04 \\
 & ~~~~~~~~~~''  & 2012 Oct 14 & E & 13.6 & 19.57$\pm$0.01 				& 0.91$\pm$0.02 & 0.58$\pm$0.01 & -- & 1.49$\pm$0.01 \\
 & ~~~~~~~~~~''  & 2013 Oct 01 & E & 2.7 & 20.64$\pm$0.01 	& 0.85$\pm$0.03 & 0.54$\pm$0.02  & -- & 1.39$\pm$0.03 \\
 & ~~~~~~~~~~''   & 2013 Oct 01 & E & 5.4 & 20.23$\pm$0.01 	& 0.94$\pm$0.03 & 0.51$\pm$0.02  & -- & 1.45$\pm$0.03 \\
 & ~~~~~~~~~~''   & 2013 Oct 01 & E& 8.2 & 19.96$\pm$0.01 	& 0.95$\pm$0.03 & 0.50$\pm$0.02  & -- & 1.45$\pm$0.03 \\
 & ~~~~~~~~~~''   & 2013 Oct 01 & E & 10.8 & 19.80$\pm$0.01 	& 0.91$\pm$0.03 & 0.52$\pm$0.02  & -- & 1.43$\pm$0.03 \\

14 & C/2013 C2  & 2014 Feb 26 & E & 2.7 & 19.51$\pm$0.02 & 0.87$\pm$0.02 & 0.55$\pm$0.02 & -- & 1.42$\pm$0.02 \\
 & ~~~~~~~~~~''  & 2014 Feb 26 & E & 5.4 & 19.14$\pm$0.02 & 0.88$\pm$0.02 & 0.54$\pm$0.02 & -- & 1.42$\pm$0.02 \\
 & ~~~~~~~~~~''  & 2014 Feb 26 & E & 8.0 & 18.91$\pm$0.02 & 0.90$\pm$0.02 & 0.52$\pm$0.02 & -- & 1.42$\pm$0.02 \\
 & ~~~~~~~~~~''  & 2014 Feb 26 & E & 10.8 & 18.77$\pm$0.02 & 0.90$\pm$0.02 & 0.52$\pm$0.02 & -- & 1.42$\pm$0.02 \\

15 & C/2013 P4    & 2014 Oct 22 & E  &  2.7  & 19.78$\pm$0.02 & 0.82$\pm$0.02 & 0.49$\pm$0.02 & -- & 1.31$\pm$0.02 \\
 & ~~~~~~~~~~''          &            &     E &                    2.7 & 19.25$\pm$0.02& 0.83$\pm$0.02 &0.48$\pm$0.02 & -- & 1.31$\pm$0.02 \\
 & ~~~~~~~~~~''            &         &     E &                    4.0 & 18.99$\pm$0.02 & 0.83$\pm$0.02 & 0.49$\pm$0.02 & -- & 1.32$\pm$0.02 \\
 
16 & 166P/NEAT &                 2015 Feb 17 & E & 2.7 & 23.16$\pm$0.02 & 0.89$\pm$0.11 & 0.56$\pm$0.03 & -- & 1.45$\pm$0.11 \\
17 & 167P/CINEOS & 2010 Sep 10 & E & 8.2 & 20.89$\pm$0.03 			& 0.80$\pm$0.04 & 0.57$\pm$0.03  & 0.45$\pm$0.03 & 1.37$\pm$0.04\\
 & ~~~~~~~~~~''  & 2012 Oct 13          & E & 13.6 & 21.21$\pm$0.03 			& 0.68$\pm$0.05 & 0.34$\pm$0.05 & -- & 1.12$\pm$0.04 \\
                   
\\

 & Sun\tablenotemark{d} & -- & -- & -- & -- & 0.64$\pm$0.02 & 0.35$\pm$0.01 & 0.33$\pm$0.01 & 0.99$\pm$0.02 \\

\enddata

\tablenotetext{a}{Morphology; P = point source, E = extended}
\tablenotetext{b}{Angular diameter of photometry aperture in arcsecond}
\tablenotetext{c}{Apparent red magnitude measured in aperture of diameter $\phi$}
\tablenotetext{d}{Holmberg et al.~2006}

\end{deluxetable}
\end{landscape}



\clearpage


\begin{landscape}
\begin{deluxetable}{llllll}
\tabletypesize{\scriptsize}
\tablecaption{Color Comparison\tablenotemark{a, b}
\label{summary}}
\tablewidth{0pt}
\tablehead{
\colhead{Object}    & \colhead{$B-V$} & \colhead{$V-R$} & \colhead{$R-I$}  
& \colhead{$B-R$}  & \colhead{Source} }
\startdata
Cold Classical KBOs (Cold CKBO) & 1.09, 1.06$\pm$0.02(13) & 0.64, 0.66$\pm$0.02(13) & N/A & 1.73, 1.72$\pm$0.02(13) & Tegler online\\
Mean KBO Colors & 0.93, 0.92$\pm$0.02(85) & 0.57, 0.57$\pm$0.02(85) & N/A & 1.52, 1.49$\pm$0.03(85) 	& Tegler online\\
Plutinos       & 0.94, 0.93$\pm$0.04(25) & 0.58, 0.56$\pm$0.03(25) & N/A& 1.52, 1.49$\pm$0.06(25) 		& Tegler online\\
Inactive Centaurs & 0.85, 0.93$\pm$0.04(29) & 0.51, 0.55$\pm$0.03(29) & 0.44, 0.45$\pm$0.02(3) & 1.30, 1.47$\pm$0.06(32) & Table (\ref{centaur_photometry})  + Peixinho et al.~(2003, 2012) \\
Hot Classical KBOs (Hot CKBO) & 0.95, 0.89$\pm$0.05(14) & 0.57, 0.54$\pm$0.04(14) & N/A & 1.52, 1.44$\pm$0.08(14) & Tegler online\\
Scattered KBOs (SKBO) & 0.82, 0.84$\pm$0.03(20) & 0.54, 0.53$\pm$0.02(20) & N/A & 1.38, 1.37$\pm$0.05(20) & Tegler online\\

JFC Nuclei & 0.80, 0.87$\pm$0.05(16) & 0.49, 0.50$\pm$0.03(16) & 0.47, 0.46$\pm$0.03(12) & 1.31, 1.37$\pm$0.08(12) & Lamy and Toth~(2009) \\
Damocloids & 0.79, 0.80$\pm$0.02(20) & 0.51, 0.51$\pm$0.02(21) & 0.48, 0.47$\pm$0.02(15) & 1.31, 1.31$\pm$0.02(20) & Table (\ref{damo_photometry}) + Jewitt (2005) \\
Active Centaurs & 0.80, 0.80$\pm$0.03(12) & 0.51, 0.50$\pm$0.03(13) & 0.57, 0.57$\pm$0.03(8) & 1.29, 1.30$\pm$0.05(12) & Table (\ref{centaur_photometry}) + Jewitt (2009) \\
Active LPC & 0.77, 0.78$\pm$0.02(25) & 0.46, 0.47$\pm$0.02(24) & 0.41, 0.42$\pm$0.03(7) & 1.23, 1.24$\pm$0.02(25) & Table (\ref{lpc_photometry})\\
Active JFC & 0.74, 0.75$\pm$0.02(26)   & 0.46, 0.47$\pm$0.02(26) & 0.44, 0.43$\pm$0.02(26) & 1.21, 1.22$\pm$0.02(26) & Solontoi et al.~(2012)\\
LPC Nuclei & 0.76, 0.77$\pm$0.02(5) & 0.45, 0.44$\pm$0.02(5) & 0.43, 0.44$\pm$0.02(5) & 1.22, 1.22$\pm$0.03(5) & Lamy and Toth (2009) \\
Jupiter Trojans\tablenotemark{c} & 0.73$\pm$0.02($\sim$1000) & 0.45$\pm$0.02($\sim$1000) & 0.43$\pm$0.02($\sim$1000) & 1.18$\pm$0.02($\sim$1000)&  Szabo et al.~(2007) \\
\hline
Sun  & 0.64$\pm$0.02 & 0.35$\pm$0.01 & 0.33$\pm$0.01 & 0.99$\pm$0.02 & Holmberg et al.~(2006)\\

\enddata




\tablenotetext{a}{For each object group  we list the median color, the mean color with its $\pm$1$\sigma$ standard error, and the number of measurements}

\tablenotetext{b}{Ordered by mean B-R color}

\tablenotetext{c}{The  Trojan  data are presented in such a way that we cannot determine the median or error on the mean directly; we set the latter equal to $\pm$0.02 magnitudes to reflect likely systematic errors in transforming from the Sloan filter system to BVRI. }

\end{deluxetable}
\end{landscape}

\clearpage

\begin{deluxetable}{lllll}
\tablecaption{B-R vs.~Aperture Radius\tablenotemark{a}
\label{color_gradient}}
\tablewidth{0pt}
\tablehead{
\colhead{Object}    & \colhead{$\phi$=1.35\arcsec} & \colhead{$\phi$=2.70\arcsec} & \colhead{$\phi$=4.05\arcsec}  
& \colhead{$\phi$=5.40\arcsec}  }
\startdata

C/2006 S3 & 1.08$\pm$0.02 & 1.13$\pm$0.02 & 1.16$\pm$0.02 & 1.16$\pm$0.02 \\
C/2012 Q1 & 1.39$\pm$0.05   & 1.45$\pm$0.01  & 1.45$\pm$0.01  & 1.43$\pm$0.01   \\
C/2013 C2 & 1.42$\pm$0.01 & 1.42$\pm$0.01 & 1.42$\pm$0.01 & 1.42$\pm$0.01  \\
C/2014 W6 &  1.20$\pm$0.04 & 1.23$\pm$0.04 & 1.25$\pm$0.04 & 1.21$\pm$0.04 \\
C/2014 XB8 & 1.21$\pm$0.03 & 1.20$\pm$0.03 &1.14$\pm$0.03 & 1.12$\pm$0.03 \\
C/2014 R1 & 1.24$\pm$0.01 & 1.27$\pm$0.01 & 1.27$\pm$0.01 & 1.28$\pm$0.01 \\
C/2015 B2 & 1.21$\pm$0.02 & 1.24$\pm$0.02 & 1.23$\pm$0.02 & 1.24$\pm$0.02 \\

\enddata




\tablenotetext{a}{We list the mean B-R color with its $\pm$1$\sigma$ standard error.  The columns show the angular radii of the circular apertures used to take the measurements, in arcseconds.}

\end{deluxetable}


\clearpage

\clearpage

\clearpage

\clearpage

\clearpage

\clearpage

\clearpage

\clearpage

\clearpage

\clearpage

\clearpage

\clearpage

\clearpage

\clearpage

\clearpage

\clearpage

\end{document}